\documentclass[amsmath,twocolumn,superscriptaddress,prb,aps]{revtex4-1} 
\usepackage{amsthm,amsfonts,graphicx,verbatim, color}
\usepackage{bm}
\usepackage{hyperref}
\usepackage{hhline}

\newcommand{\be}{\begin{equation}}
\newcommand{\ee}{\end{equation}}
\newcommand{\bea}{\begin{eqnarray}}
\newcommand{\eea}{\end{eqnarray}}

\newcommand{\Tr}{{\rm Tr}\,}
\renewcommand{\epsilon}{\varepsilon}
\renewcommand{\vec}[1]{{\bf #1}}

\newcommand{\bq}{{\vec{q}}}

\newcommand\dvec{{\boldsymbol{d}}}

\newcommand{\diag}{\mathrm{diag}}

\def\beq{\begin{equation}}
\def\eeq{\end{equation}}
\def\bea{\begin{eqnarray}}
\def\eea{\end{eqnarray}}

\begin{document}

\title{Disorder-driven destruction of a non-Fermi liquid semimetal via renormalization group}
\author{Rahul M. Nandkishore}
\affiliation{Department of Physics and Center for Theory of Quantum Matter, University of Colorado at Boulder, Boulder, CO 80309, USA}
\author{S. A. Parameswaran}
\affiliation{Department of Physics and Astronomy, University of California, Irvine, CA 92697, USA}

\begin{abstract}
We investigate the interplay of Coulomb interactions and short-range-correlated disorder in three dimensional systems where absent disorder the  non-interacting band structure hosts a quadratic band crossing. Though the {\it clean} Coulomb  problem is believed to host a `non-Fermi liquid' phase, disorder and Coulomb interactions have the same scaling dimension in a renormalization group (RG) sense, and thus should be treated on an equal footing. We therefore implement a controlled $\epsilon$-expansion and apply it at leading order to derive RG flow equations valid when disorder and interactions are both weak.  
We find that the `non-Fermi liquid' fixed point is  unstable to disorder, and demonstrate that the problem inevitably flows to strong coupling, outside the regime of applicability of the perturbative RG. An examination of the flow to strong coupling suggests that disorder is asymptotically more important than interactions, so that the low energy behavior of the system can be described by a {\it non-interacting} sigma model in the appropriate symmetry class (which depends on whether exact particle-hole symmetry is imposed on the problem). We close with a discussion of general principles unveiled by our analysis that dictate the interplay of disorder and Coulomb interactions in gapless semiconductors, and of connections to many-body localized systems with long-range interactions.
\end{abstract}
\maketitle
\section{Introduction}

Semimetals, gapless semiconductors with properties intermediate between metals and insulators, provide a fascinating playground for exploration of condensed matter physics. The past decade his witnessed an explosion of activity studying 
such semimetals (for a recent review, see Ref.~\onlinecite{VafekVishwanath}). Interest has focused for the most part on two dimensional systems such as graphene~\cite{Graphenereview1, Graphenereview2} and topological insulator surface states \cite{TI1, TI2}, but in recent years has broadened also to include three dimensional systems such as  Weyl and Dirac semimetals  \cite{MurakamiWeyl, WanSavrasov, PKV, Moll, SonSpivak, ParameswaranGroverAbanin, anomaly1, anomaly2, fradkin, NandkishoreHuseSondhi, pixleyhuse}. A particularly rich problem, first examined by Abrikosov in 1971, involves three dimensional systems with {\it quadratic} band crossings and Coulomb interactions, argued to be in a   {\it non-Fermi liquid}  phase. 
 Interest in this problem has recently been revived~\cite{MoonXuKimBalents, Herbut, LABIrridate} because of its relevance for pyrochlore iridates. However, theoretical explorations of this unusual three dimensional system have largely been confined to the {\it clean} (disorder-free) problem, whereas realistic materials are always disordered to some degree.

The interplay of disorder and interactions has separately generated enormous theoretical and experimental activity, catalyzed most recently by a surge of interest in many-body localization (MBL)~\cite{Anderson1958, gmp, baa, Imbrie, MBLARCMP}. While research into MBL has focused primarily on systems at high (even infinite) temperatures, the combined role of disorder and interactions at {\it zero} temperature is also worthy of study. Three dimensional quadratic band crossings are a particularly promising setting for exploring the {\it zero temperature} interplay of disorder and interactions, because both short range correlated disorder and Coulomb interactions are {\it relevant} with the {\it same} scaling dimension, indicating that they should be treated on an equal footing. 

Here, we investigate this interplay of 
 interactions and disorder at zero temperature in three dimensional materials hosting quadratic band crossings.  We use a renormalization-group (RG) procedure to analyze the  scaling behavior of weak short-range-correlated disorder and Coulomb interactions, working with the most general (symmetry constrained) Hamiltonian for these systems. 
 Unlike earlier work~\cite{LaiRoyGoswami} that employed an uncontrolled truncation of perturbation theory, we use a controlled $\epsilon$-expansion RG scheme 
that treats  disorder and interactions on an equal footing. We derive the perturbative RG equations to $O(\epsilon)$, and demonstrate that the Abrikosov fixed point is unstable to disorder. We further demonstrate that there are no stable fixed points within the domain of validity of the perturbative RG, and the problem flows to strong coupling. An analysis of the flow to strong coupling reveals that disorder is asymptotically {\it more} important than Coulomb interactions, so 
 that the problem at strong coupling should admit a {\it non-interacting} sigma model description. We argue that at strong coupling there should exist a {\it localized} phase, even though the bare Hamiltonian contains power-law long range interactions. We discuss how this result  interfaces with the claimed obstructions to MBL in higher dimensions and with long range interactions \cite{avalanches, Burin, yaodipoles}. We conclude with a discussion of general principles revealed by our analysis. 

The remainder of this paper is structured as follows. We begin in Sec.~\ref{sec:symm} with a  discussion of the basic symmetry-constrained action, and the RG scheme that will be employed to analyze it. In Sec.~\ref{sec:betafunc} we derive the RG flow equations to $O(\epsilon)$, and analyze these in Sec.~\ref{sec:RGanalysis}. We demonstrate that there are no stable fixed points within the domain of applicability of perturbative RG, and that the Abrikosov fixed point is unstable to disorder. Sec.~\ref{sec:strongflow} examines the flow to strong coupling, and the likely behavior of the system in this limit. We close with a discussion of general principles revealed by the analysis, as well as the connections to recent developments in MBL in Sec.~\ref{sec:conclusions}. 

\section{\label{sec:symm}Preliminaries}
\subsection{Symmetry-constrained action}
For three-dimensional quadratic band crossings, the low energy bands form 
 a four-dimensional representation of the lattice symmetry group \cite{MoonXuKimBalents}, and the $\vec{k} \cdot \vec{p}$ Hamiltonian for the clean non-interacting system takes the form \cite{Herbut}
 \begin{equation}
 \mathcal{H}_0 = \sum_{a=1}^N \frac{d_a(\vec{k}) }{2m} \Gamma^a \label{bare},
 \end{equation}
where the $\Gamma_a$ are the {rank four irreducible representations of the Clifford algebra relation $\{\Gamma_a,\Gamma_b\} = 2\delta_{ab}1$, where $\{A,B\} = AB+BA$ is the anticommutator. There are $N=5$ such matrices, which are related to the familiar gamma matrices from the Dirac equation (plus the matrix conventionally denoted as $\gamma_5$), but are not quite identical because we are working with Euclidean metric $\{\Gamma_a,\Gamma_b\} = 2\delta_{ab}1$ instead of Minkowski metric $\{\Gamma_a,\Gamma_b\} = 2\times\diag(-1,+1,+1,+1)$. Throughout, we will make extensive use of the defining Clifford algebra relation, and the fact that $\sum_a \Gamma_a\Gamma_a =N$. }
Meanwhile the  
$d_a(k)$ are  $l=2$, spherical harmonics of the form
\begin{eqnarray}\label{ddef}
d_1(\vec{k}) &=& \sqrt{3} k_y k_z; d_2(\vec{k}) =  \sqrt{3} k_x k_z; d_3(\vec{k}) =  \sqrt{3} k_x k_y; \nonumber\\ d_4(\vec{k})&=&\frac{\sqrt{3} }{2} (k_x^2 - k_y^2), \quad d_5(\vec{k}) = \frac{1}{2} (2 k_z^2 - k_x^2 - k_y^2)
\end{eqnarray}
A lattice system is allowed to have 
 anisotropy terms in the $\vec{k} \cdot \vec{p}$ Hamiltonian, reflecting the reduced rotational symmetry, and also an isotropic $\frac{k^2}{2m'}$ term with no spinor structure. These additional terms were shown to be irrelevant in the RG sense in the presence of Coulomb interactions \cite{Abrikosov, MoonXuKimBalents} and we therefore ignore them here. In principle it is worth revisiting the irrelevance of these anisotropy terms in the presence of disorder, but this is beyond the scope of the present work.  
 Henceforth, we restrict our considerations to the idealized Hamiltonian (\ref{bare}), which contains the `most relevant' terms in the $\vec{k} \cdot \vec{p}$ Hamiltonian of the clean system. 
 
  {We note in passing that the 3D quadratic band crossing problem is very different, both qualitatively and in detail, to the 3D `Schrodinger' problem of free fermions in the continuum, which also have an $E\sim k^2$ dispersion. In the Schrodinger problem, the quadratic dispersion arises when the chemical potential is placed at the bottom of the band i.e. when the system is prepared at {\it zero density}. This problem is effectively single particle in nature. In contrast, in the 3D quadratic band crossing considered here, the chemical potential is placed at the intersection of two bands, and this is a truly many body problem exhibiting phenomena such as screening, which are an integral part of the analysis. Additionally, of course, the Dirac structure of the problem quantitatively effects the detailed structure of the loop corrections.}
 
We now discuss the symmetry properties of the Hamiltonian. Note that of the five $\Gamma$ matrices, three can be chosen pure real (e.g. $\Gamma_{1,2,3}$) and two pure imaginary ($\Gamma_{4,5}$). The  Hamiltonian then has a time reversal symmetry with a time reversal operator that can be represented~\cite{Herbut} as $T = \Gamma_4 \Gamma_5 K$, where $K$ is complex conjugation; we see that $T^2=-1$.

We wish to examine the interplay of Coulomb interactions and disorder in these quadratic band crossing materials, treating disorder using a replica field theory. The most general   Euclidean time action with these properties may be written $S = S_0 + S_{s} + S_{v}$, where %
\begin{eqnarray}
S_0 &=& \sum_{i=1}^n \int d\tau d^dx \left[\psi_i^{\dag} [\partial_{\tau} - \mathcal{H}_0 + i e \varphi_i ] \psi_i + \frac{c}{2} (\nabla \varphi_i)^2  \right]\,\,\,\,\,\,\,\,
\end{eqnarray}
describes a clean quadratic semimetal with Coulomb interactions propagated by a scalar boson field $\varphi$, and
   \begin{eqnarray}
   S_{s} &=& - \sum_{i,j=1}^n W_0  \int d\tau d\tau' d^d x (\psi^{\dag}_i \psi_i)_{\tau} (\psi^{\dag}_j \psi_j)_{\tau'},   \nonumber\\ 
S_{v} &=& - \sum_{M,N,i,j}  W_{MN}\int d\tau d\tau' d^d x (\psi^{\dag}_i M \psi_i)_{\tau} (\psi^{\dag}_j N \psi_j)_{\tau'}. \label{action1}
\end{eqnarray}
are terms representing short-range-correlated disorder with and without spinor structure (which we refer to as scalar and vector disorder respectively), parametrized by constants $W$. We treat disorder  in the replica formalism with replica indices $i, j$; as is usual, we will take the number of replicas $n\rightarrow 0$ at the end of our computation. The sums over $M$ and $N$  range over all independent $4\times 4$ non-identity Hermitian matrices in the spinor space. 
We note that we have neglected short range interactions in $S_0$ since these are less relevant in an RG sense than either long-range (Coulomb) interactions or short-range-correlated disorder. 

Reasonable assumptions on disorder and the use of symmetries allow a considerable simplification of the vector disorder term. First, assuming that the different components of vector disorder are independent we may fix $W_{MN} = W_M \delta_{MN}$, so that
\begin{eqnarray}
S_{v} &=& - \sum_{M,i,j}  W_{M}\int d\tau d\tau' d^d x (\psi^{\dag}_i M \psi_i)_{\tau} (\psi^{\dag}_j M \psi_j)_{\tau'}. \label{action2}
\end{eqnarray}
This may be further simplified by noting that the sum over $M$ in (\ref{action2}) ranges over {\it all} possible (non-identity) Hermitian rank four matrices. In $d=3$, the space of $4\times 4$ Hermitian matrices is spanned by the identity matrix, the five $4\times 4$ Gamma matrices $\Gamma_a$ and the ten matrices $\Gamma_{ab} = \frac{1}{2i} [\Gamma_a, \Gamma_b]$. Now, rotations in the spinor space transform the $\Gamma_a$ into one another (and likewise the $\Gamma_{ab}$), and so if we assume that disorder respects the isotropy of the problem in spinor space, then in $d=3$ we have only two independent vector disorder parameters 
\begin{eqnarray}
S_{v} &=& - W_1 \sum_{(a)} 
\sum_{ij=1}^n  \int d\tau d\tau' d^d x (\psi^{\dag}_i \Gamma_a \psi_i)_{\tau} (\psi^{\dag}_j \Gamma_a \psi_j)_{\tau'}\\& & - W_2 \sum_{(ab)}
 \sum_{ij=1}^n  \int d\tau d\tau' d^d x (\psi^{\dag}_i \Gamma_{ab} \psi_i)_{\tau} (\psi^{\dag}_j \Gamma_{ab} \psi_j)_{\tau'} \nonumber
\,\,\,\,\end{eqnarray}
[Henceforth we will explicitly show summations over replica indices, but not over  spatial indices of $\Gamma$-matrices; from now, we will adopt the Einstein convention for the latter, i.e. repeated indices are summed.]
We now note that the matrices $\Gamma_{ab}$ are odd under time reversal, so disorder of this form {\it locally} breaks time reversal symmetry \cite{Herbut}, even while preserving time reversal symmetry on average. We will impose exact time reversal symmetry, thereby setting $W_2=0$ henceforth (extending the analysis to incorporate $W_2 \neq 0$ would be an interesting avenue for future work). It will further be convenient to treat Coulomb interactions and disorder on an equal footing, and therefore we integrate out the scalar boson $\phi$ to obtain the Coulomb interaction as an effective four fermion term. This then yields a final action 
\begin{widetext}
\begin{eqnarray}
S &=& \sum_{i=1}^n \left[\int d\tau d^d x\, \psi_i^{\dag} [\partial_{\tau} - \mathcal{H}_0 ] \psi_i  + \frac{e^2}{2 c } \int d\tau d^d q\, d^d p\, d^d p'\, 
V(q)\psi^{\dag}_{\vec{p},i} \psi^{\dag}_{\vec{p'},i} \psi_{\vec{p'-q},i} \psi_{\vec{p+q},i}\right] \nonumber \\ &-& W_0 \sum_{i,j=1}^n   \int d\tau d\tau' d^d x (\psi^{\dag}_i \psi_i)_{\tau} (\psi^{\dag}_j \psi_j)_{\tau'} 
- W_1 \sum_{i,j=1}^n  \int d\tau d\tau' d^d x (\psi^{\dag}_i \Gamma_a \psi_i)_{\tau} (\psi^{\dag}_j \Gamma_a \psi_j)_{\tau'}, \label{fullactionMoon}\end{eqnarray}
\end{widetext}
where the Coulomb interaction $V(q) = \frac{1}{q^2}$  has been written in momentum space.  
 
We may also consider particle-hole symmetry, implemented~\cite{Boettcher} by taking $\Gamma \rightarrow - \Gamma^*$ and simultaneously complex conjugating the Hamiltonian. If the $\Gamma$-matrices are written in the Weyl basis, this can be implemented by taking~\footnote{http://eduardo.physics.illinois.edu/phys582/582-chapter7.pdf} $C = i\sigma_1 \otimes \sigma_2 K$, that manifestly squares to $-1$. The $W_0$ term locally breaks particle-hole symmetry, and thus if we demand that disorder locally preserve particle-hole symmetry then we can further set $W_0=0$. 
We shall discuss the RG flows both in the presence and in the absence of particle-hole symmetry.
 
 If we set the scaling dimensions $[x^{-1}] = 1$ and $[\tau^{-1}] = z$, then invariance of the bare action fixes $[\psi] = d/2$, $[m]= 2-z$ and $[\varphi] = \frac{d+z-2}{2}$, where $z=2$ at tree level from (\ref{bare}). Power counting the quartic terms then gives
\begin{equation}
[e^2] = z+2-d \qquad [W_{M}] = 2z - d - 2 \eta_{M}  \label{dimensions}
\end{equation}
where we have allowed for anomalous exponents $[\psi^{\dag} \psi] = d + \eta_{0}$ and $[\psi^{\dag} \Gamma \psi] = d + \eta_{1}$ (here, we only consider $M = 0, 1$, though $W_2$ and indeed all the $W_{MN}$ have similar tree-level scaling). The anomalous exponents are zero at the Gaussian fixed point. Note that Coulomb interactions and disorder are both relevant at tree level with the {\it same} exponent, at least at the Gaussian fixed point about which we will be perturbing, and so must be treated on an equal footing. 

\subsection{Regularization schemes}
We wish to analyze the action (\ref{fullactionMoon}) using a controlled renormalization scheme that treats disorder and interactions on an equal footing. The RG scheme first employed to treat three dimensional quadratic band crossings \cite{Abrikosov, Beneslavski} involves a continuation to $d=4-\epsilon$ dimensions. In $d=4$, the Coulomb interaction is marginal at tree level, and loop corrections can therefore be computed to order $\epsilon$ (a description of the physical situation requires a continuation to $\epsilon = 1$, which could be problematic \cite{Herbut}). At first glance, disorder is also marginal at tree level in $d=4$, so dimensional continuation would seem appropriate. However a straightforward application of the Abrikosov dimensional continuation technique is not suitable for the disordered problem, for 
 reasons that we now describe. 

A straightforward generalization of (\ref{bare}) to four spatial dimensions extends the theory to include $N=9$ gamma matrices, each of which is rank {\it sixteen}. Thus (\ref{bare}) in $d=4$ is defined in terms of %
$16$-component spinor fields, and hence with %
$16\times 16$ matrices $M$ in (\ref{action2}). In $d=4$ the gamma matrices and their commutators do not provide a complete basis for the space of Hermitian rank-$16$ %
 matrices, and so the most general action containing short range disorder in $d=4$ contains {\it unphysical} disorder types with no analog in the $d=3$  problem of interest.  Additionally, while the theory in $d=4$ still has a time reversal symmetry (TRS), the time reversal operator squares~\cite{Herbut} to $+1$; therefore, analytic continuation in dimensionality changes the representation of time reversal, potentially crucially altering the disorder physics. [A dramatic example of the sensitivity of disorder to the representation of TRS  is the existence of a localization transition in  $d=2$ disordered spin-orbit coupled systems in the symplectic class with $T^2=-1$, and its absence for spin-rotationally invariant systems where $T^2=+1$ that are always localized.]

An alternative regularization scheme developed by Moon {\it et al} \cite{MoonXuKimBalents} involves continuing to four dimensions while assuming that the angular and gamma matrix structure remains as in $d=3$ i.e. radial momentum integrals are performed with respect to a $d=4-\epsilon$ dimensional measure $\int \frac{p^{3-\epsilon} dp}{(2\pi)^{4-\epsilon}}$, but angular momentum integrals are performed only over the three dimensional sphere parametrized by polar and azimuthal angles $\theta$ and $\varphi$. Nevertheless, the overall angular integral of an angle-independent function $\int_{\hat{\Omega}}\cdot 1$ is taken to be $2\pi^2$ (as is appropriate for the total solid angle in  $d=4$), and angular integrals are normalized accordingly . Therfore, the angular integrations are performed with respect to the measure 
\be\label{moonmeasure}
\int dS (\ldots) \equiv \frac{\pi}{2} \int_0^{\pi} d \theta \int_0^{2\pi} d \varphi \sin \theta (\ldots),\ee
 where the $\pi/2$ is inserted for the sake of normalization. We refer to this as the Moon scheme and employ it henceforth in our analysis.

\section{\label{sec:betafunc}Computation of RG Equations}
\subsection{Non-Fermi liquid in the clean system} 

We begin by developing the analysis for the clean system, setting $W_0=0$ and $W_1=0$. This allows us to compare to existing results in the literature. Loop corrections will all be calculated in $d=4$, whereupon there will be log divergences in integrals instead of the $\sim 1/\epsilon$ scaling that will obtain at non-zero $\epsilon$. Any log divergence should therefore be understood as a shorthand for $1/\epsilon$.  Note that an alternative is to compute the loop integrals directly in $d=4-\epsilon$, however a potential drawback with this approach is that it can on occasion yield spurious $O(\epsilon)$ contributions to $\beta$-functions, in situations where a naive log divergence is rendered finite owing to a nontrivial angular dependence of the integrand.

\begin{figure}
a) \includegraphics[width = 0.4 \columnwidth]{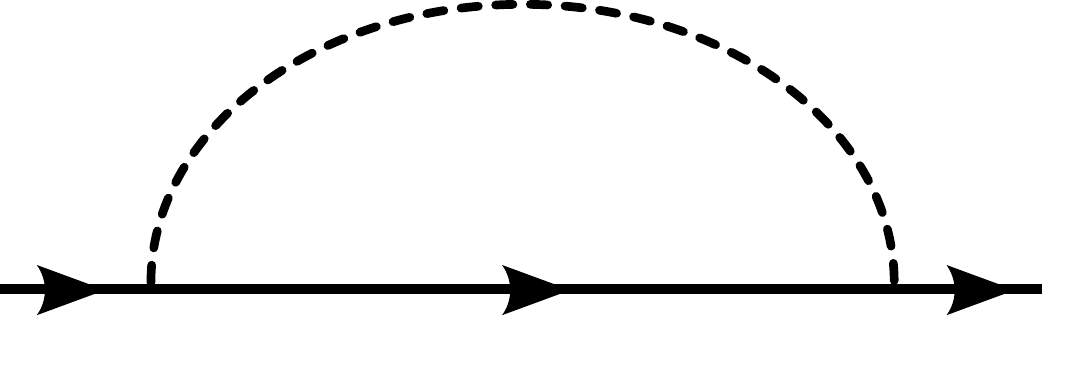}
b) \includegraphics[width = 0.4\columnwidth]{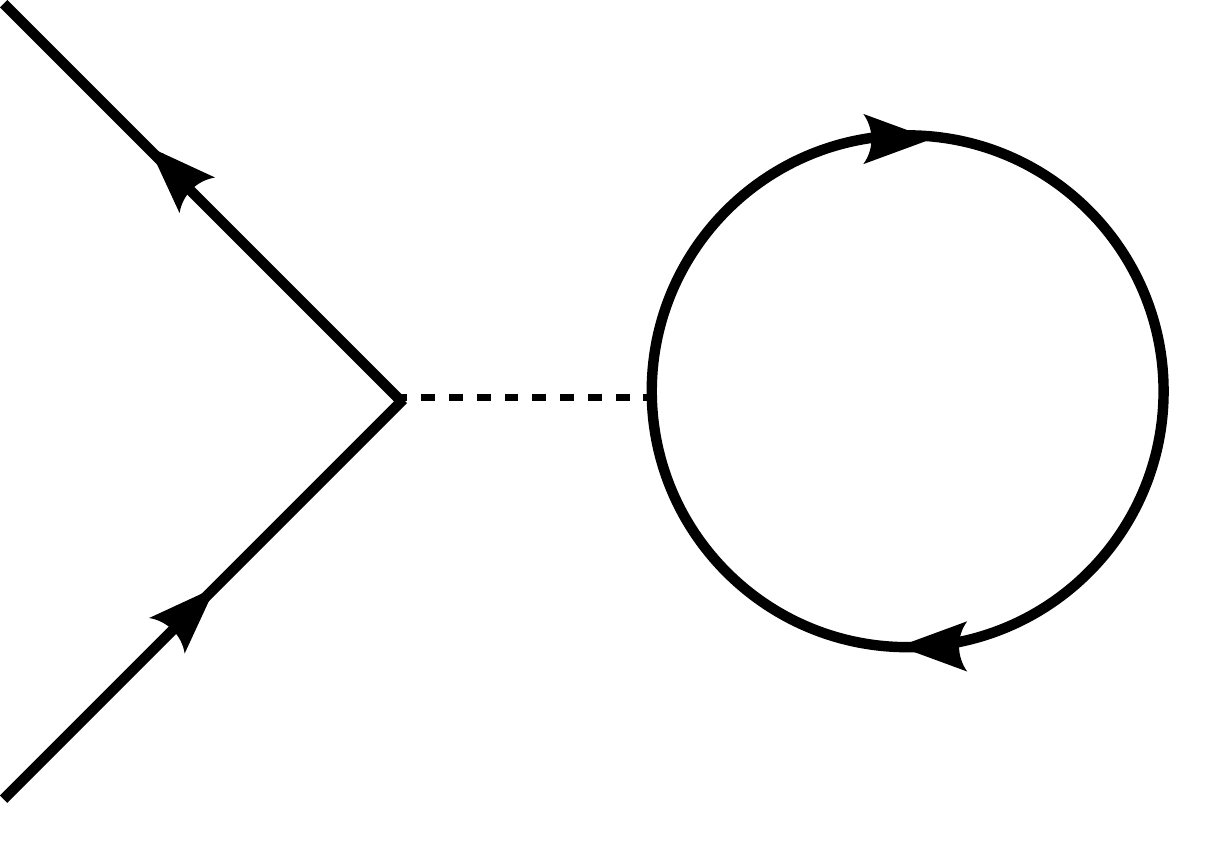}
\caption{\label{scalardisorderonlygf} The two diagrams shown above determine the $O(\epsilon)$ correction to the Green function. Solid lines represent the bare Green function. The dashed line may represent either disorder or the Coulomb interaction. If the dashed line represents disorder, then it connects two fermion lines at the same point in real space, but the two fermions may have different time indices and replica indices. In this case the diagram (b) is proportional to $n$, the number of replica flavors, and vanishes upon taking the replica limit $n \rightarrow 0$.  If the dashed line represents the Coulomb interaction, then it connects two fermions with the same time index and same replica index, but with different spatial position. }
\end{figure}

For the clean system we can drop the replica indices, since the interaction is replica-diagonal.  
 We work perturbatively, expanding the path integrals in powers of $e^2/2c$, assumed small. At leading order, we obtain 
\be
\frac{e^2}{2c} \int{\frac{d^dp\, d^dp'\, d^dq}{(2\pi)^{3d}}}\, V(q) \psi^{\dag}_{\vec{p}} \psi^{\dag}_{\vec{p'}} \psi_{\vec{p'-q}} \psi_{\vec{p+q}}.
\ee   
There are two possible contractions  (matching one $\psi^{\dag}$ with one $\psi$) of this term that renormalize  
the electron Green's function (see Fig.~\ref{scalardisorderonlygf}). 
`Hartree' contractions of the real space form $\langle \psi^{\dag}_x \psi_x \rangle \psi^{\dag}_{x'} \psi_{x'}$ correspond to tadpole diagrams, which simply shift the overall chemical potential and can be ignored (we assume the renormalized chemical potential is at the quadratic band crossing point), but `exchange'  contractions of the form $\langle \psi^{\dag}_x \psi_{x'} \rangle \psi^{\dag}_{x'} \psi_x$ cannot be ignored. Upon re-exponentiating, these correct the electron Green's function, via $G^{-1} = G_0^{-1} - \Sigma$, where $G_0^{-1}$ is the bare Green's function and $ \Sigma(\omega, \vec{k}) =- 2 \frac{e^2}{2c} \int_q V(q) G_0(\Omega + \omega, \vec{k} + \vec{q})$ is the self energy, and the combinatorial factor of $2$ is because there are the two possible exchange contractions.  
Since the bare Green's function can be written as 
\begin{equation}
G_0(\omega, \vec{k}) =  \frac{i \omega + \dvec(\vec{k}) \cdot{\boldsymbol{\Gamma}}}{\omega^2 +|\dvec(\vec{k})|^2} \label{baregf}
\end{equation}
where $|d(\vec{k})|^2 = (\frac{k^2}{2m})^2$, the self energy takes the form
\begin{equation}
\Sigma(\vec{k}) \!=\! -\frac{e^2}{c} \!\int_{-\infty}^{\infty}\! \frac{d\Omega}{2\pi}\!\! \int\!\! \frac{d^d q}{(2\pi)^d} \frac{i (\omega + \Omega) + \dvec(\vec{k} + \vec{q}) \cdot{\boldsymbol{\Gamma}}}{(\omega+\Omega)^2 + |\dvec(\vec{k} + \vec{q})|^2} V(q). 
\end{equation}
Shifting $\Omega \rightarrow \omega + \Omega$ removes the dependence on $\omega$; performing the remaining integral over $\Omega$ by the method of residues and then shifting $\vec{k+q} \rightarrow \vec{q}$, we obtain 
\begin{equation}
\Sigma(\vec{k}) = -\frac{e^2}{2c}  \int \frac{d^d q}{(2\pi)^d} \frac{\dvec_{\vec{q}} \cdot{\boldsymbol{\Gamma}}}{|\dvec_{\vec{q}}|} V(|\vec{q} - \vec{k}|) 
\end{equation}
The $k=0$ component of this just renormalizes the chemical potential (in fact, this precisely cancels the contribution of the `tadpole' diagrams, which have a relative minus sign due to the fermion loop). Subtracting off the $k=0$ piece, we therefore obtain a self energy 
\begin{equation}
\Sigma(\vec{k}) =- \frac{e^2}{2c}  \int \frac{d^d q}{(2\pi)^d} \frac{\dvec_{\vec{q}} \cdot{\boldsymbol{\Gamma}}}{|\dvec_{\vec{q}}|} [V(|\vec{q} - \vec{k}|) - V(q)].
\end{equation}
In the spirit of the RG, the `internal' momenta $q$ should be taken to be much larger than the `external' momenta $k$  (any divergences coming from $q < k$ are spurious). We can therefore expand the term in square brackets in powers of $k/q$ and thus evaluate the integral: 
\begin{eqnarray}
\Sigma(\vec{k}) &=& -\frac{e^2}{2c} \int \frac{q^3 dq }{(2\pi)^4}   \times\\& &  \int dS\, \hat{\dvec}_{\vec{\bq}}%
\cdot{\boldsymbol{\Gamma}} \frac{1}{q^2}\left[2 \frac{k}{q}  \cos \theta - \frac{k^2}{q^2} + \frac{4 k^2 \cos^2 \theta}{q^2}\right]\nonumber
\end{eqnarray}
where we choose $\vec{k}$ to lie along the $z$ axis without loss of generality, $dS$ represents  the angular measure (\ref{moonmeasure}) in the Moon scheme, and $\hat{\dvec}_{\vec{q}} = \frac{\dvec(\vec{q})}{|\dvec(\vec{q})|}$ is an angular function independent of the magnitude of $\vec{k}$, that may be computed from (\ref{ddef}).
With $\vec{k}$ chosen to lie along the $z$ axis, the angular integral vanishes for $\Gamma_{1-4}$ and we obtain 
\begin{eqnarray}
\Sigma(\vec{k})&=& -\frac{e^2}{2c} \frac{\pi}{2} k^2 \Gamma_5 \int \frac{q^3 dq }{(2\pi)^3 q^4} \int_0^{\pi} \sin\theta d\theta \nonumber\\ & &\times \frac{1}{2} (2 \cos^2 \theta - \sin^2 \theta) (- 1 + 4 \cos^2 \theta),
\end{eqnarray}
where the factor of $\pi/2$ is from the Moon scheme (\ref{moonmeasure}). 
The integrals may now be performed (the radial integral is log divergent in UV and IR and needs to be regulated), and yields, after a little algebra,
\begin{equation}
\Sigma(\vec{k}) = -\frac{m e^2}{15 \pi^2 c}  \log \frac{\Lambda_{\text{UV}}}{\Lambda_{\text{IR}}} \dvec(\vec{k}) \cdot{\boldsymbol{\Gamma}} .
\end{equation}
We have made use of rotational symmetry to note that even though the self energy is proportional to $d_5 \Gamma_5$ for $\vec{k}$ along the $z$ axis, the full self energy for general $\vec{k}$ must be of the form $\dvec \cdot \vec{\Gamma}$ (rearranging in this form also generates the overall factor of $m$). %
In order to determine the RG flows, it is convenient to return to the spirit of the momentum-shell approach and take $\Lambda_{\text{IR}} = \Lambda_{\text{UV}}e^{-l}$ where $l$ is the RG flow parameter, assumed small: in other words, we only consider internal momenta within a shell near the cutoff. In this fashion, we find that the one-loop renormalized Green's function becomes
\begin{equation}
G^{-1} = i\omega - \dvec(\vec{k}) \cdot{\boldsymbol{\Gamma}} \left(1 - \frac{m e^2}{15 \pi^2 c} l\right).
\end{equation}
From this, we see that the mass  has scaling dimension $[m] = 2 - z - \frac{m e^2}{15 \pi^2 c}$, so that in order that it is invariant under RG we must alter the dynamic exponent to 
\begin{equation}
z = 2  - \frac{m e^2}{15 \pi^2 c}, 
\end{equation}
in agreement with Ref.~\onlinecite{MoonXuKimBalents}. 

\begin{figure*}
(a) \includegraphics[width = 0.45 \columnwidth]{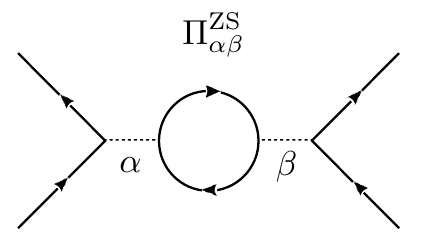}
(b) \includegraphics[width = 0.45 \columnwidth]{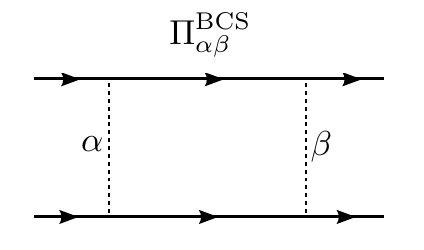}
(c) \includegraphics[width = 0.45 \columnwidth]{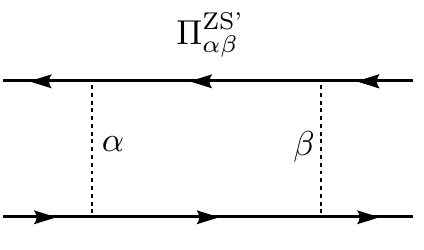}
(d) \includegraphics[width = 0.45 \columnwidth]{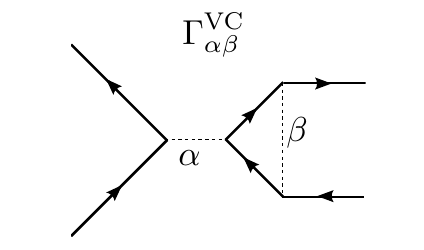}
\caption{\label{scalardisordervertices} Diagrams that determine the $O(\epsilon)$ correction to the four-fermion vertices (either disorder or interaction). These diagrams may be denoted respectively as ZS, BCS, ZS' and VC, following the naming convention from Ref.~\onlinecite{Shankar}. Solid lines denote the fermion Green function. Dashed lines may represent either disorder ($\alpha,\beta = 0, 1$) or the Coulomb interaction ($\alpha, \beta =c$). If the dashed line represents disorder, then it connects two fermion lines at the same point in real space, but the two fermions may have different time indices and replica indices.  If the dashed line represents the Coulomb interaction, then it connects two fermions with the same time index and same replica index, but with different spatial position. Note that unlike the ZS, BCS, and ZS' diagrams,  the VC diagrams are generically not symmetric under interchange of indices, i.e. $\Gamma^{\text{VC}}_{\alpha\beta}\neq\Gamma^{\text{VC}}_{\beta\alpha}$.  } 
\end{figure*}

At the next order in the expansion of the action, we generate a term (in real space) of the form 
\begin{align}\label{Coulombcontraction} \left(\frac{e^2}{2c}\right)^2&\int d\tau dx dx' dx'' dx'''\, \frac12 \left[\left(\psi^{\dag}_x \psi_x \psi^{\dag}_{x'} \psi_{x'}\right) \right.\\&
\left.\times\left(\psi^{\dag}_{x''} \psi_{x''} \psi^{\dag}_{x'''} \psi_{x'''}\right) \frac{1}{|x-x'||x''-x'''|}\right]\nonumber.\end{align}
 Performing two contractions, we obtain a correction to the Coulomb interaction. This can be represented diagrammatically;  as usual, only `fully connected' diagrams contribute, which can be labelled, in the standard terminology of Ref.~\onlinecite{Shankar}, as ZS, BCS, and ZS', as well as a vertex correction (VC) (Fig.\ref{scalardisordervertices}). 

The ZS diagram corresponds to contracting fermion lines in the `bubble' topology. The four distinct ways to do this lead to an overall combinatorial factor of four, and the  fermion loop adds a factor of $-N_f$ ($N_f$ is the number of fermion flavors), resulting in a correction to the action of the form
\be
\delta S = \int\frac{d^dp\,d^dp'\,d^dq}{(2\pi)^{3d}}\, \Pi^{\text{ZS}}_{\text{cc}}(\bq) \psi^{\dag}_{p} \psi^{\dag}_{p'} \psi_{p'+q} \psi_{p-q},\ee 
where 
\begin{equation}
\Pi^{\text{ZS}}_{\text{cc}}(\bq)= - \frac{4 N_f}{2 q^4}\left(\frac{e^2}{2c}\right)^2\!\!\!\Tr\!\!\! \int\frac{d\omega}{2\pi}\frac{dk}{(2\pi)^d}G\left(\omega, \vec{k}+ {\vec{q}}\right) G\left(\omega, \vec{k} \right),
\end{equation}
and the trace (denoted $\Tr$) is over the spinor indices. We use the subscript `cc' to indicate that this diagram emerges from contractions of the product of two Coulomb terms, and the ZS labels the diagram topology. 
Using the form of the Green's function (\ref{baregf}), we obtain
\begin{eqnarray}
\Pi^{\text{ZS}}_{\text{cc}}(\bq) &=& - \frac{2 N_f}{q^4}\left(\frac{e^2}{2c}\right)^2\int \frac{d^dk}{(2\pi)^d}\\
& & \times \Tr\int \frac{d\omega}{2\pi} \frac{(i\omega + \dvec(\vec{k+q}) \cdot{\boldsymbol{\Gamma}})(i\omega + \dvec(\vec{k}) \cdot{\boldsymbol{\Gamma}})}{(\omega^2 + |d(\vec{k+q})|^2)(\omega^2 + |d(\vec{k})|^2)}\nonumber.
\end{eqnarray}
Performing the $\omega$ integral by the method of residues, and dropping terms that will manifestly vanish upon performing the angular integral, we find
\begin{align}
\Pi^{\text{ZS}}_{\text{cc}}(\bq)= -  \frac{rN_f}{q^4} &\left(\frac{e^2}{2 c}\right)^2  \int \frac{d^dk}{(2\pi)^d} \frac{1}{|\dvec(\vec{k+q})| + |\dvec(\vec{k})|}\nonumber\\&\times \left( \frac{\dvec(\vec{k+q}) \cdot \dvec(\vec{k})}{|\dvec(\vec{k+q})||\dvec(\vec{k})|}-1\right),
\end{align}
where $r\equiv \Tr \mathbf{1} =4$ is the dimension of the $\Gamma$ matrices. The above integral manifestly vanishes for $k=0$. Taylor expanding in small $k$, we obtain
\begin{eqnarray}
\Pi^{\text{ZS}}_{\text{cc}}(\bq) &=&  -  \frac{rN_f}{q^4} \left(\frac{e^2}{2 c}\right)^2 \!\!\!
 \int \frac{d^dk}{(2\pi)^d} \frac{2m }{|\vec{k+q}|^2 + |\vec{k}|^2}  
 \frac{3q^2 }{2 k^2} \sin^2 \!\theta
 \nonumber\\
&=& -  \frac{rN_f}{q^4} \left(\frac{e^2}{2 c}\right)^2\!\!\!\int \frac{d^dk}{(2\pi)^d} \frac{m}{ k^2} \times \frac{3q^2 }{2 k^2} \sin^2\! \theta,
\end{eqnarray}
where again we have taken $\vec{q}$ to lie along the $z$ axis. Performing angular integrals in the Moon scheme (where $r=4$) we find  \begin{eqnarray}
\Pi^{\text{ZS}}_{\text{cc}}(\bq)  &=&   \frac{N_f} {q^2}\left(\frac{e^2}{2 c}\right)^2  \frac{m}{2\pi^2}   \int_{\Lambda_{\text{IR}}}^{\Lambda_{\text{UV}}} k^3dk \frac{1}{ k^4} \nonumber\\ &=&  \frac{N_f} {q^2}\left(\frac{e^2}{2 c}\right)^2  \frac{m}{2\pi^2} \log \frac{\Lambda_{\text{UV}}}{\Lambda_{\text{IR}}}.\label{PiZScc}
\end{eqnarray}
Re-exponentiating and again using ${\Lambda_{\text{IR}}} = {\Lambda_{\text{UV}}}e^{-l}$  we find that this term renormalizes the Coulomb interaction, changing its coefficient via 
$\frac{e^2}{2 c} \rightarrow \frac{e^2}{2 c}   \left[ 1 - \frac{m e^2}{4 \pi^2 c} N_f  l \right]$. As we show in Appendix~\ref{sec:CleanRGOnlyZS}, the remaining diagram topologies (VC, ZS', BCS) do not contribute to the clean-system RG flows. Defining a dimensionless interaction parameter $u = \frac{m e^2}{8 \pi^2 c}$ we obtain the RG equations 
\begin{eqnarray}
z &=& 2 - \frac{8}{15} u\\ \frac{du}{dl} &=& (z+2-d) u - 2 N_f u^2  = \epsilon u - \frac{30 N_f +8}{15}u^2 
\end{eqnarray}
where we used $d = 4-\epsilon$ and incorporated feedback from the dispersion renormalization. These are the same equations as Ref.~\onlinecite{MoonXuKimBalents}, and have two fixed points.
First, there is a trivial (Gaussian) fixed point at $u=0$ with $z=2$, which is unstable, and a (stable) non-Fermi liquid fixed point, where%
\begin{equation}
u_* =\frac{15}{30 N_f + 8} \epsilon; \qquad z = 2 - \frac{4}{15 N_f + 4} \epsilon.
\end{equation}
It is the fate of this non-Fermi liquid fixed point upon the addition of scalar and vector disorder that is our central interest in the balance of this paper.

\subsection{The disordered non-interacting problem}
\label{puredisorder}
We now introduce disorder. Before considering (\ref{fullactionMoon}) in its entirety, we begin by determining the RG flow equations in the presence of disorder alone, and only then turn to the interplay of disorder and Coulomb interactions. As before, all calculations are done directly in $d=4$, using the Moon regularization scheme for angles. 

We therefore begin by considering (\ref{fullactionMoon}) with $V(q)$ set to zero, and work perturbatively in weak disorder, as parametrized by $W_0$ and $W_1$. At leading order, we obtain diagrams renormalizing the Green's function which are shown in Fig.\ref{scalardisorderonlygf}. Upon taking the replica limit $n \rightarrow 0$ we obtain a self energy that is diagonal in the replica space and takes the form (repeated indices are summed)
\begin{align}
\Sigma(\omega, \vec{k})   &= {2}W_0 \int \frac{d^d p}{(2\pi)^d} \, G(\omega, \vec{p}) \nonumber\\&\,\,\,\,\,+2W_1 \int \frac{d^d p}{(2\pi)^d} \,\Gamma_{a} G(\omega, \vec{p}) \Gamma_{a}\nonumber \\&=  ( W_0 + N  W_1) \frac{ im^2\omega}{\pi^2} \log\frac{\Lambda_{\text{UV}}}{\Lambda_{\text{IR}}}
\end{align}
where the relative minus sign compared to the Coulomb diagram reflects the relative sign between the two terms in the action. We have  performed angular integrals with the Moon scheme, and used the fact that an integral of an isolated $l=2$ harmonic vanishes, i.e. $\int dS \hat{d_a}(\vec{k}) = 0$. Re-exponentiating, we obtain a Green's function
\begin{align}
G^{-1} &= i \omega \left( 1 -  \frac{  m^2 (W_0 + N  W_1)}{\pi^2} \log\frac{\Lambda_{\text{UV}}}{\Lambda_{\text{IR}}} \right) - \dvec(\vec{k}) \cdot{\boldsymbol{\Gamma}}\nonumber\\&\approx \left( 1 - \frac{m^2 (W_0 + N  W_1)}{\pi^2} \log\frac{\Lambda_{\text{UV}}}{\Lambda_{\text{IR}}} \right)\nonumber\\&\,\,\,\,\,\times \left[i \omega - \bigg( 1 +  \frac{m^2(W_0 + N  W_1)}{\pi^2} \log \bigg) \dvec(\vec{k}) \cdot{\boldsymbol{\Gamma}} \right]
\end{align}
where the last approximation is valid in the perturbative regime, where $W_{0,1} \log\frac{\Lambda_{\text{UV}}}{\Lambda_{\text{IR}}} < 1$. 
This implies that the quasiparticle residue $Z$ renormalizes according to \be
\frac{d Z^{-1}}{dl} = -\frac{m^2( W_0+N  W_1)}{ \pi^2}.
\ee
We also infer that the mass has scaling dimension 
\be
[m] = 2 - z + \frac{m^2(W_0+NW_1)}{ \pi^2},
\ee
so that requiring its invariance under the RG yields a dynamical exponent  
\begin{equation}
z = 2 + \frac{m^2 }{ \pi^2 } (W_0 + N  W_1).
\end{equation}

We now turn to the loop corrections to the disorder lines themselves. These come from the fully connected contractions of 
{
\begin{align}\label{disordercontraction}
\delta S = &\frac12 \int d\tau d\tau'  d\tau'' d\tau'''d^d x  d^d x'\\& \times \sum_{i,j,k,l=1}^n\left[ W_0^2 (\psi^{\dag}_i \psi_i)_{x}^{\tau} (\psi^{\dag}_j \psi_j)_{x}^{\tau'} 
(\psi^{\dag}_k \psi_k)_{x'}^{\tau''} (\psi^{\dag}_l \psi_l)_{x'}^{\tau'''} \right.
\nonumber \\ &+ W_1^2 (\psi^{\dag}_i \Gamma^i_{a} \psi_i)_{x}^{\tau} (\psi^{\dag}_j \Gamma^j_{a} \psi_j)_{x}^{\tau'} 
(\psi^{\dag}_k \Gamma^k_{b} \psi_k)_{x'}^{\tau''} (\psi^{\dag}_l \Gamma^l_{b} \psi_l)_{x'}^{\tau'''}\nonumber\\
&+ \left. 2 W_0 W_1  (\psi^{\dag}_i \psi_i)_{x}^{\tau} (\psi^{\dag}_j \psi_j)_{x}^{\tau'} 
(\psi^{\dag}_k \Gamma^k_{b} \psi_k)_{x'}^{\tau''} (\psi^{\dag}_l \Gamma^l_{b} \psi_l)_{x'}^{\tau'''} 
\right]\nonumber
\end{align}}
where repeated $\Gamma$-matrix indices are as usual summed over, and {we have kept track of the replica indices on $\Gamma$ matrices}. As before, the contractions lead to four distinct diagram topologies (Fig.~\ref{scalardisordervertices}), that we will denote by the same labels as the one-loop corrections to the interaction in the clean case. We now discuss each of these in turn. 
 In doing so it is useful to recall that while $\Gamma$ matrices with the same 
 {replica} index anticommute, $\Gamma$ matrices with different 
{replica indices act on independent spaces} and hence commute. The vertex that a given one-loop diagram renormalizes can be read off from its final $\Gamma$-matrix structure; diagrams with no $\Gamma$-matrices are to be understood as proportional to identity in the spinor space.  We will  continue to sum over repeated indices, even when not explicitly stated. %

\noindent\underline{\textbf{(i) ZS.}} This diagram comes with a factor of $n$ (the number of replica indices) and thus vanishes upon taking the replica limit $n \rightarrow 0$.

\noindent\underline{\textbf{(ii) VC.}} This has an overall minus sign relative to ZS because of the absence of a fermion loop; the results (summarized in Table~\ref{VC}) depend on the number of scalar and vector disorder lines that enter the diagram. 
 A VC diagram with two scalar ($W_0$) lines can emerge in $8$ distinct ways; including factors from (\ref{disordercontraction}) we find a correction to the scalar vertex from
 \begin{align}
 \Gamma^{\text{VC}}_{00} &=  4 W_0^2%
 \int\frac{d^d k}{(2\pi)^d} \frac{(i\omega + \dvec_\vec{k}\cdot{\boldsymbol{\Gamma}}^{{j}}) (i\omega + \dvec_\vec{k} \cdot{\boldsymbol{\Gamma}}^{{j}})}{(\omega^2 + k^4/(2m)^2)^2}\nonumber\\&= 4 W_0^2   %
 \int\frac{d^d k}{(2\pi)^d} \frac{ -\omega^2 + k^4/(2m)^2}{(\omega^2 + k^4/(2m)^2)^2}
 \nonumber\\&=\frac{(2m)^2 }{2 \pi^2}W_0^2\log \frac{\Lambda_{\text{UV}}}{\Lambda_{\text{IR}}}\label{GammaVC00} ,
 \end{align}
where we have used $d_{a}(\vec{k}) =  k^2\hat{d}_a(\vec{k})$, %
$\int dS\, \hat{d}_{a}(\vec{k}) \hat{d}_{b}(\vec{k}) = {2\pi^2} \delta_{ab}/N $
and $|\dvec(\vec{k})|^2 = k^{4}/(2m)^2$ to simplify the numerator and drop terms that vanish upon angular integration. In the final step, we have used the fact that disorder is static, so that $\omega$ is an external frequency that can be set to zero to evaluate the diagram, yielding the logarithmic correction. We note parenthetically that, if we were to take $\omega\neq 0$, it may serve as the IR cutoff (a similar result holds for all the other diagrams discussed in this section). However, as we are interested in the situation where internal momenta are near the UV cutoff, we may take the IR cutoff to be much greater than $\omega$ and thereby set $\omega$ to zero safely.  Returning to the scalar-scalar VC diagram and  re-exponentiating (with the appropriate fields for the external legs included) we find that this makes a correction to the scalar disorder term of the form $\delta W_0 = + \frac{2 m^2 }{ \pi^2} W_0^2 l$. 

A VC diagram with two vector ($W_1$) lines may also emerge from 8 distinct contractions, and after  
{setting the external frequency $\omega=0$} and simplifying leads to
 \begin{align}
\frac{\Gamma^{\text{VC}}_{11}}{(2m)^2} &=  4  W_1^2 
\Gamma_{a}^{{i}} \int\frac{d^d k}{(2\pi)^d} \frac{\Gamma_{b}^{{j}} [\hat{\dvec}_{\vec{k}} \cdot{\boldsymbol{\Gamma}}^{{j}}] \Gamma_{a}^{{j}} [\hat{\dvec}_{\vec{k}} \cdot{\boldsymbol{\Gamma}}^{{j}}] \Gamma_{b}^{{j}}}{k^4}\nonumber\\
 &=  \frac{W_1^2}{2 N \pi^2}  \times \Gamma_{a}^{{i}}  \Gamma_{b}^{{j}} \Gamma_{c}^{{j}} \Gamma_{a}^{{j}} \Gamma_{c}^{{j}} \Gamma_{b}^{{j}}\times \log \frac{\Lambda_{\text{UV}}}{\Lambda_{\text{IR}}} \nonumber\\
 &=- \frac{(N-2)  W_1^2}{2 N \pi^2}    
 \Gamma_{a}^{{i}} \Gamma_{b}^{{j}}  \Gamma_{a}^{{j}}  \Gamma_{b}^{{j}} \times l\nonumber\\
 &= \frac{(N-2)^2  W_1^2}{2 N \pi^2}    
 \Gamma_{a}^{{i}}  \Gamma_{a}^{{j}}   \times l.\end{align}
 where we have relied on similar identities as in $\Gamma^{\text{VC}}_{00}$ in completing the angular integrals and used ${\Lambda_{\text{IR}}}= {\Lambda_{\text{UV}}}e^{-l}$.
 Note the index structure of the RHS, indicating that this corrects the vector disorder term.
 Inserting  fields for the external legs of this diagram and re-exponentiating, we find that $\delta W_1 = \frac{2 (N-2)^2m^2}{N\pi^2}W_1^2 l$.
 
Finally,   VC diagrams with mixed $W_0$ and $W_1$ lines can emerge in one of two ways, each of which corrects a different bare vertex and comes with a combinatorial factor of $4$. Including the factors from (\ref{disordercontraction}), we find that the scalar vertex receives a contribution from 
\begin{align}
\frac{\Gamma^{\text{VC}}_{01}}{(2m)^2} &=4  W_0 W_1 \int\frac{d^dk}{(2\pi)^d}\frac{\Gamma_{b}^{{j}} [\hat{\dvec}_{\vec{k}} \cdot{\boldsymbol{\Gamma}}^{{j}}]  [\hat{\dvec}_{\vec{k}}\cdot{\boldsymbol{\Gamma}}^{{j}}] \Gamma_{b}^{{j}}}{{k^4} }\nonumber\\&= \frac{W_0 W_1} {2 N \pi^2} \times\Gamma_{b}^{{j}} \Gamma_{c}^{{j}} \Gamma_{c}^{{j}} \Gamma_{b}^{{j}}\times \log\frac {\Lambda_{\text{UV}}}{\Lambda_{\text{IR}}} 
 \nonumber\\&= \frac{NW_0 W_1}{2 \pi^2} \times l, \label{GammaVC01} \end{align}
 that, following previous examples, yields  $\delta W_0 =  \frac{2Nm^2}{\pi^2} W_0  W_1 l$. Similarly, the vector disorder vertex is corrected via  \begin{align}
\frac{\Gamma^{\text{VC}}_{10}}{(2m)^2}&=4 W_1 W_0 \Gamma_{\mu}^{{i}} \int\frac{d^dk}{(2\pi)^d} \frac{[\hat{\dvec}_{\vec{k}} \cdot{\boldsymbol{\Gamma}}^{{j}}] \Gamma_{a}^{{j}} [\hat{\dvec}_{\vec{k}} \cdot{\boldsymbol{\Gamma}}^{{j}}]}{k^4} \nonumber\\&= \frac{W_1 W_0}{2N \pi^2}  \Gamma_{a}^{{i}} {\log}\frac{\Lambda_{\text{UV}}}{\Lambda_{\text{IR}}} 
\Gamma_{b}^{{j}}  \Gamma_{a}^{{j}} \Gamma_{b}^{{j}}\nonumber\\&= -  \frac{(N-2)W_1 W_0}{2 N \pi^2}
\times l\times \Gamma^{{i}}_{a} \Gamma^{{j}}_{a},
\end{align}
and renormalizes the vector disorder term as $\delta W_1 = - \frac{2(N-2)m^2}{{N \pi^2}} W_0 W_1 l.$

\begin{table}
\begin{tabular}{|c|c|c||c|}
\hline
Coupling  & $ \lambda_0$ & $ \lambda_1 $ & $u$ \tabularnewline
\hline
$ \lambda_0$ & $\delta \lambda_0 = \lambda_0^2l $ & $\delta \lambda_1 = -   \frac{N-2}{N } \lambda_0 \lambda_1l $& $\delta u = \lambda_0 ul $ \tabularnewline
\hline
$ \lambda_1$ &$\delta \lambda_0 =  N\lambda_0 \lambda_1l$&$\delta \lambda_1 =  \frac{(N-2)^2}{ N} \lambda_1^2 l$& $ \delta u =  N\lambda_1 ul$ \tabularnewline
\hhline{|=|=|=||=|}
$u$ &0 &$\delta  \lambda_1 =  2\frac{N-1}{N }  \lambda_1 u  l$&0\tabularnewline
\hline
\end{tabular}
\caption{Contributions to the $\beta$-functions from the VC diagrams. Here, $\lambda_i = \frac{2m^2W_i}{\pi^2}$, $u = \frac{me^2}{8 \pi^2 c}$, and $l$ is the RG flow parameter. \label{VC}} \end{table}

\noindent\underline{\textbf{(iii) BCS and ZS':}}  
Each of these one-loop corrections (summarized in Table~\ref{BCS}) comes with a combinatorial factor of $4$, and depends on the nature of the internal disorder lines. 
Scalar-scalar BCS and ZS' diagrams give identical contributions of the  form
\begin{align}
\frac{\Pi_{00}^{\text{BCS}}}{(2m)^2} =\frac{\Pi_{00}^{\text{ZS'}}}{(2m)^2} &= 2 W_0^2 \int\frac{d^d k}{(2\pi)^d} \frac{[\hat\dvec_\vec{k}\cdot{\boldsymbol{\Gamma}}^{{i}}][ \hat\dvec_\vec{k} \cdot{\boldsymbol{\Gamma}}^{{j}}]}{k^4} \nonumber\\&= 2 W_0^2  \Gamma_{a}^{{i}} \Gamma_{b}^{{j}} \int\frac{d^d k}{(2\pi)^d}  \frac{\hat{d}_a(\vec{k})\hat{d}_b(\vec{k})}{k^4} \nonumber\\&=  \frac{W_0^2}{4 N\pi^2}  \Gamma_{a}^{{i}} \Gamma_{a}^{{j}}  \log \frac{\Lambda_{\text{UV}}}{\Lambda_{\text{IR}}}.
\end{align}
Adding the two together and reexponentiating gives a correction   $ \delta W_1 = + \frac{2m^2 }{N \pi^2} W_0^2 l$ to the vector disorder term. 

In the vector-vector case, the BCS diagram gives \begin{align}
\frac{\Pi_{11}^{\text{BCS}}}{(2m)^2} &=
2 W_1^2\int \frac{d^dk}{(2\pi)^d} \frac{\left[ \Gamma_{a}^{{i}} \hat{\dvec}_\vec{k} \cdot \vec{\Gamma}^{{i}} \Gamma_{b}^{{i}} \right] \left[ \Gamma_{a}^{{j}} \hat{\dvec}_\vec{k} \cdot \vec{\Gamma}^{{j}} \Gamma_{b}^{{j}} \right]}{k^4}
\end{align}
whereas the ZS' diagram has different gamma-matrix index structure,
\begin{align}
\frac{\Pi_{11}^{\text{ZS'}}}{(2m)^2} &=
2 W_1^2\int \frac{d^dk}{(2\pi)^d} \frac{\left[ \Gamma_{a}^{{i}} \hat{\dvec}_\vec{k} \cdot \vec{\Gamma}^{{i}} \Gamma_{b}^{{i}} \right] \left[ \Gamma_{b}^{{j}} \hat{\dvec}_\vec{k} \cdot \vec{\Gamma}^{{j}} \Gamma_{a}^{{j}} \right]}{k^4}
\end{align}
Adding these together and simplifying we find
\begin{align}
\frac{\Pi_{11}^{\text{BCS}}}{(2m)^2} +\frac{\Pi_{11}^{\text{ZS'}}}{(2m)^2} &=
\frac{W_1^2}{4N\pi^2}\log \frac{\Lambda_{\text{UV}}}{\Lambda_{\text{IR}}} \Gamma_a^{{i}}\Gamma_{c}^{{i}}\Gamma_{b}^{{i}} \nonumber\\&\,\,\,\,\,\,\,\,\,\,\,\,\,\,\times \left(\Gamma_a^{{j}}\Gamma_{c}^{{j}}\Gamma_{b}^{{j}}+\Gamma_b^{{j}}\Gamma_{c}^{{j}}\Gamma_{a}^{{j}}\right)\nonumber\\&= \frac{W_1^2(3N-2)}{2N\pi^2}\Gamma_a^{{i}}\Gamma_a^{{j}}\times l,
\end{align}
leading to a correction to the scalar disorder term of the form $\delta W_1 = \frac{ (3N-2)m^2}{2N\pi^2} W_1^2l$.

Finally turning to the vector-scalar cross terms, we find 
\begin{align}
\frac{\Pi_{01}^{\text{BCS}}}{(2m)^2} +\frac{\Pi_{01}^{\text{ZS'}}}{(2m)^2} &= 4W_0W_1\int \frac{d^dk}{(2\pi)^d} \frac{\left[ \Gamma_{a}^{{i}} \hat{\dvec}_\vec{k} \cdot \vec{\Gamma}^{{i}} \right]}{k^4}\nonumber\\ 
&\,\,\,\,\,\,\,\,\,\,\,\,  \left[ \Gamma_{a}^{{j}} \hat{\dvec}_\vec{k} \cdot \vec{\Gamma}^{{j}} + \hat{\dvec}_\vec{k} \cdot \vec{\Gamma}^{{j}}\Gamma_{a}^{{j}} \right]\nonumber\\
&=\frac{W_0W_1}{2N\pi^2}\log \frac{\Lambda_{\text{UV}}}{\Lambda_{\text{IR}}} \Gamma_{a}^{{i}}\Gamma_b^{{i}} \{\Gamma_a^{{j}},\Gamma_b^{{j}} \}\nonumber\\ 
&= \frac{W_0W_1}{\pi^2} l,
\end{align}
thereby correcting the scalar disorder term via $\delta W_0 = \frac{4m^2}{\pi^2} W_0 W_1 l$.

\begin{table}
\begin{tabular}{|c|c|c||c|}
\hline
Coupling  & $ \lambda_0$ & $ \lambda_1 $ & $u$ \tabularnewline
\hline
$ \lambda_0$ &$\lambda_1 = \frac{1}{N} \lambda_0^2 l$& $\delta \lambda_0 = 2 \lambda_0\lambda_1l$& $ 0$ \tabularnewline
\hline
$\lambda_1$ &0&$ \delta \lambda_1 =   \frac{3N-2}{N}\lambda_1^2 l$ & $0$ \tabularnewline
\hhline{|=|=|=||=|}
$u$ &0&0& $ 0$ \tabularnewline
\hline
\end{tabular}
\caption{Sum of contributions to the $\beta$-functions from the BCS and ZS' diagrams, with same conventions as Table~\ref{VC}. \label{BCS}}
\end{table} 
 
 Combining all our results, defining rescaled~\footnote{Observe that since we have fixed $[m]=0$,  $\lambda_i$ has the same scaling dimension as  $W_i$.} disorder variables 
 \begin{equation}
{ \lambda_i = \frac{(2m)^2}{2\pi^2}W_i,}
 \end{equation}
  and recalling that the tree-level scaling dimension of the disorder term is $(2z-d)$, where 
\begin{align}
z 	&= 2 + \frac{1}{2}\left(\lambda_0 + N \lambda_1\right)
 \end{align}
 is the dynamical exponent, we find the flow equations
  \begin{align}
 \frac{d \lambda_0}{dl} &= \lambda_0 \left[\epsilon  + 2\lambda_0 + 2(N+1)\lambda_1\right] \nonumber\\
 \frac{d \lambda_1}{dl} &=\epsilon\lambda_1 + \left[\frac{\lambda_0^2}{N} + \frac{2N^2-N+2}{N} \lambda_1^2  +\frac{2}{N}\lambda_0\lambda_1\right]
  \end{align}
where we recalled that $\epsilon = 4-d$. Recall also that the initial disorder couplings are non-negative. Now we can readily verify that both $\beta$ functions are strictly positive except at $\lambda_0=0=\lambda_1$. In other words, the  trivial Gaussian fixed point is unstable and the system flows to strong disorder: the non-interacting disordered problem does not have any stable fixed points at weak disorder, within the regime of applicability of a perturbative calculation.

\subsection{Coulomb interactions and disorder}

We finally turn to the full problem, including both Coulomb interactions and disorder. Our first step is to examine the correction to the electron Green's function; at leading order it may be obtained simply by combining the contributions of interactions and disorder acting separately. This leads us to conclude that, at this order, the quasiparticle residue renormalizes according to
\begin{equation}
\frac{d Z^{-1}}{dl}=  -\frac{\lambda_0 + N  \lambda_1}{2},
\end{equation}
and that the dynamical exponent must be
\begin{equation}
z = 2  + \frac{\lambda_0 + N  \lambda_1}{2}  -   \frac{8}{15}u \label{eq:fullanomdim}
\end{equation}
in order that the mass remain invariant under the RG. 
Meanwhile, the four-fermion pieces of the action --- both the Coulomb interaction and the replicated disorder terms --- acquire an additional renormalization from fully connected contractions of 
\begin{align}
\delta S &= - \frac{e^2}{2c}\int d\tau d\tau'  d\tau'' d^dx d^d x' d^d x''\sum_{i,j,k=1}^n  \label{coulombplusdisordercontraction}\\
& \,\,\,\,\,\,\,\,\,\frac{(\psi^{\dag}_k \psi_k)_{\tau'', x'} (\psi^{\dag}_k \psi_k)_{\tau'', x''}}{ |x'-x''|^2} 
\left[W_0 (\psi^{\dag}_i \psi_i)_{\tau, x} (\psi^{\dag}_j \psi_j)_{\tau', x}\right.  \nonumber\\
&\,\,\,\,\,\,\,\,\,\,\,\,\,\,\,\,\,\,\,\,\,\,\,\,\,\,\,\,\,\,\,\,\,\,\,\, \,\,\,\,   \left.+ W_1 (\psi^{\dag}_i \Gamma^{{i}}_{a} \psi_i)_{\tau, x} (\psi^{\dag}_j \Gamma^{{j}}_{a} \psi_j)_{\tau', x}\right].\nonumber
\end{align}
Note the absence of an overall factor of $\frac12$ in this expression compared with (\ref{disordercontraction}). Once gain the fully connected contractions may be labeled  ZS, VC, ZS', and BCS, but now involve one Coulomb line and one disorder line. We discuss the new contributions to the RG equations from each of these `mixed' diagrams in turn, labeling the corresponding bubbles using a similar convention as before.

\noindent\underline{\textbf{(i) ZS:}} The ZS diagram with one Coulomb line and one $W_1$ line attached vanishes upon tracing over spinor indices, $\Pi^{\text{ZS}}_{c1} = 0$. Meanwhile, the ZS diagram with one Coulomb line and one $W_0$ line attached is nearly identical to the Coulomb-only ZS diagram, except that one factor of $\frac{e^2}{2c q^2}$ (where $q$ is external momentum) is replaced by $W_0$, the sign is reversed owing to the relative sign between Coulomb and disorder terms and there is an overall factor of $2$ owing to the combinatorics of swapping Coulomb and disorder lines [Note that these differences are evident when comparing (\ref{Coulombcontraction}), (\ref{disordercontraction}), and (\ref{coulombplusdisordercontraction}).]
From the external legs of such a contraction we determine that it corrects the scalar disorder term, and from the discussion above and (\ref{PiZScc}) we see that it evaluates to 
\be
\Pi^{\text{ZS}}_{c0} = -{2 W_0}\times\frac{2cq^2}{e^2} \Pi^{\text{ZS}}_{cc} =  -N_f\frac{m e^2}{2\pi^2 c}W_0\log \frac{\Lambda_{\text{UV}}}{\Lambda_{\text{IR}}}.
\ee
Note that at leading order $\Pi^{\text{ZS}}_{c0}$ is independent of the external momentum, $\bq$; if we re-exponentiate and Fourier transform, we find that it corrects the strength of scalar disorder via $\delta W_0 = -N_f\frac{m e^2}{2\pi^2c} W_0     = -4N_f uW_0$. The additional minus sign is due to the overall negative sign in the replicated disorder term; physically, this reflects the  intuitive fact that Coulomb interactions tend to produce screening  of chemical potential fluctuations. Finally, we observe that the overall prefactor of $N_f$  stems from the assumption that the disorder coupling is `all to all' in fermion-flavor-space, just like the Coulomb interaction. A modification of this assumption would change the overall prefactor for this diagram, but we do not consider this possibility here. The result for all ZS diagrams is summarized in Table~\ref{ZS}.

\begin{table}
\begin{tabular}{|c|c|c||c|}
\hline
Coupling  & $ \lambda_0$ & $ \lambda_1 $ & $u$ \tabularnewline
\hline
$ \lambda_0$ &0& $0$& $ \delta \lambda_0 = - 4 N_f \lambda_0 u $ \tabularnewline
\hline
$\lambda_1$ &0&0 & $0$ \tabularnewline
\hhline{|=|=|=||=|}
$u$ &0&0& $\delta  u =  - 2 N_f u^2 $ \tabularnewline
\hline
\end{tabular}
\caption{Contributions to the $\beta$-functions from the ZS diagram, with same notation as in Tables~\ref{VC} and~\ref{BCS}.\label{ZS}}
\end{table}

\noindent\underline{\textbf{(ii) VC:}} We next turn to the mixed VC diagrams, that determine how the Coulomb interaction corrects a disorder vertex, and vice versa.
 The Coulomb correction to a $W_0$ disorder vertex will vanish, $\Gamma^{\text{VC}}_{0c} = 0$ just like the pure Coulomb VC diagram. However, the Coulomb correction to the $W_1$ vertex takes the form
\begin{eqnarray}
\Gamma^{\text{VC}}_{1c} &=&  - 4 W_1 \frac{e^2}{2c}\Gamma^{{i}}_{a}  \int \frac{d \omega}{2\pi} \frac{d^4 p}{(2\pi)^d}\, G(\omega, \vec{p})  \Gamma^{{j}}_{a} G(\omega, \vec{p}) \nonumber\\ &=& \frac{2e^2 W_1}{c} \Gamma^{{i}}_{a}\Gamma^{{j}}_a   \int \frac{d \omega}{2\pi} \frac{d^4 p}{(2\pi)^d}\, \frac{\omega^2 + \frac{N-2}{N} \frac{p^4}{4 m^2} }{p^{2}(\omega^2 + \frac{p^4}{4m^2})^2} \end{eqnarray}
where the combinatorial factors are similar to those for $\Gamma^{\text{VC}}_{10}$, but there is an overall minus sign because of the relative sign between (\ref{disordercontraction}) and (\ref{coulombplusdisordercontraction}), and in the second line we have commuted through the $\Gamma$ matrix as when computing  $\Gamma^{\text{VC}}_{10}$. Integrating out $\omega$ and then integrating over $p$ we obtain 
\begin{equation}
\Gamma^{\text{VC}}_{1c} =  \frac{ m W_1 e^2 }{4\pi^2c } \frac{(N-1)}{N} \Gamma^{{i}}_{a}\Gamma^{{j}}_a\frac{\Lambda_{\text{UV}}}{\Lambda_{\text{IR}}} 
\end{equation}
Upon re-exponentiating, we obtain a correction to the disorder term of form 
$\delta W_1 = W_1 (1+2 u \frac{N-1}{N }l).$ 
Note that the Coulomb term tends to {\it enhance} the vector potential disorder. At the Abrikosov fixed point, where $u^* = \frac{15}{30 N_f + 8}$, $W_1$ acquires a total vertex correction of $\frac{12}{15N_f + 4}$. Given that the term entering the action is $W_1 (\psi^{\dag} \Gamma \psi)^2$, this is equivalent to the statement that the operator $(\psi^{\dag} \Gamma \psi)$  acquires  an anomalous dimension of $-  \frac{6}{15N_f + 4}$ at the Abrikosov fixed point,  consistent with the results of Ref.~\onlinecite{MoonXuKimBalents}. 

Meanwhile, the diagrams corresponding to disorder corrections to the Coulomb vertex evaluate similarly to the disorder corrections to the $W_0$ vertex.  (i.e., to $\Gamma^{\text{VC}}_{00}$,  $\Gamma^{\text{VC}}_{01}$), except for an additional  minus sign  from (\ref{coulombplusdisordercontraction}), and the replacement of one factor of $W_0$ by $\frac{e^2}{2c q^2}$. 
For the scalar correction to the Coulomb vertex we have, using (\ref{GammaVC00}), that~\footnote{{Note a minor subtlety here: the overall combinatorial factor for $\Gamma^{\text{VC}}_{00}$ is $8$ whereas that for $\Gamma^{\text{VC}}_{c0}$ is $4$; however, the former acquires an additional factor of $\frac{1}{2}$ from (\ref{disordercontraction}), absent in (\ref{coulombplusdisordercontraction}), and hence both end up having the same overall prefactor of $4$.}}
\be
\Gamma^{\text{VC}}_{c0} = -\frac{e^2}{2c q^2 W_0}\times\Gamma^{\text{VC}}_{00} = -\frac{(2m)^2 }{2 \pi^2}\frac{e^2}{2c q^2} W_0\log \frac{\Lambda_{\text{UV}}}{\Lambda_{\text{IR}}},
\ee
whereas for the vector correction we find from (\ref{GammaVC01}) that
\be
\Gamma^{\text{VC}}_{c1} = -\frac{e^2}{2c q^2 W_0}\times\Gamma^{\text{VC}}_{01} = -\frac{(2m)^2 }{2 \pi^2} N \frac{e^2}{2c q^2} W_1\log \frac{\Lambda_{\text{UV}}}{\Lambda_{\text{IR}}}.
\ee
Re-exponentiating, we find that these shift the interaction by\begin{equation}
\delta u = \frac{(2m)^2}{2 \pi^2} (W_0+ N W_1) u l
\end{equation}
where we have picked up an additional  sign change when reincorporating the correction into the Coulomb term.

\noindent\underline{\textbf{(iii) BCS and ZS':}} Finally, we consider the ZS' and BCS diagrams. However, such diagrams with mixed disorder and Coulomb lines make vanishing contribution to the $\beta$ functions. This follows from the fact that such contractions always involve four external fermions with identical `time' index, and hence contribute solely to the renormalization of the Coulomb interaction. However, the analogous diagrams with two Coulomb lines already do not contribute to the $\beta$ function for $u$; those with mixed lines are {\it less} relevant, since there is one fewer factor of $1/p^2$ in the integrand.

 With this, we have computed all the contributions to the one-loop $\beta$ function in the full (i.e., disordered  and interacting) problem, and can now turn to analyzing the RG flows. 

\section{\label{sec:RGanalysis} Fixed Points and RG Flows}%
With the computation of the full one-loop RG structure of the theory in the previous section, we are now in a position to analyze the RG flows in the three-dimensional parameter space that includes Coulomb interactions and both scalar and vector disorder. Introducing rescaled couplings $u, \lambda_0, \lambda_1$ as before, recalling the tree-level scaling dimensions of these and collecting the results of Tables~\ref{VC}, \ref{BCS} and~\ref{ZS}, we have the one-loop flow equations
\begin{eqnarray}
\frac{d\lambda_0}{dl} &=& \lambda_0 \left[(2z-d) + \lambda_0 + (N+2) \lambda_1 - 4 N_f u \right] \\
\frac{d\lambda_1}{dl}&=&  \lambda_1 \left[ (2z-d) -\lambda_0 \frac{N-2}{N}\right.\nonumber\\
& & \left. + \lambda_1  \frac{N^2-N+2}{N} + 2 u \frac{N-1}{N} \right] + \frac1N \lambda_0^2 \\
\frac{du}{dl} &=& u \left[(z+2-d)  - 2 N_f u +  (\lambda_0 + N \lambda_1) \right].
\end{eqnarray}
Upon substituting in (\ref{eq:fullanomdim}) for the anomalous dimension 
and using $d = 4-\epsilon$, we obtain
\begin{eqnarray}
\frac{d\lambda_0}{dl}&=& \lambda_0 \left[\epsilon + 2 \lambda_0 + 2(N+1) \lambda_1 - \frac{60 N_f + 16}{15} u \right] \label{flowequations1}\\
\frac{d\lambda_1}{dl}&=&  \lambda_1 \left[ \epsilon +\lambda_0 \frac{2}{N} + \lambda_1  \frac{2 N^2-N+2}{N}\right. \nonumber\\& &\,\,\,\,\,\,\,\,\,\,\,\,\,\,\,\,\,\,\,\,\,\,\,\,\,\,\,\, \left.+ 2 u \left(\frac{N-1}{N} - \frac{8}{15} \right)\right] + \frac1N \lambda_0^2 \label{flowequations2} \\
\frac{du}{dl} &=& u\left[ \epsilon - \frac{30 N_f + 8}{15} u + \frac32 (\lambda_0 + N \lambda_1) \right]. \label{flowequations3}
\end{eqnarray}
\begin{figure}[t!]
\includegraphics[width=\columnwidth]{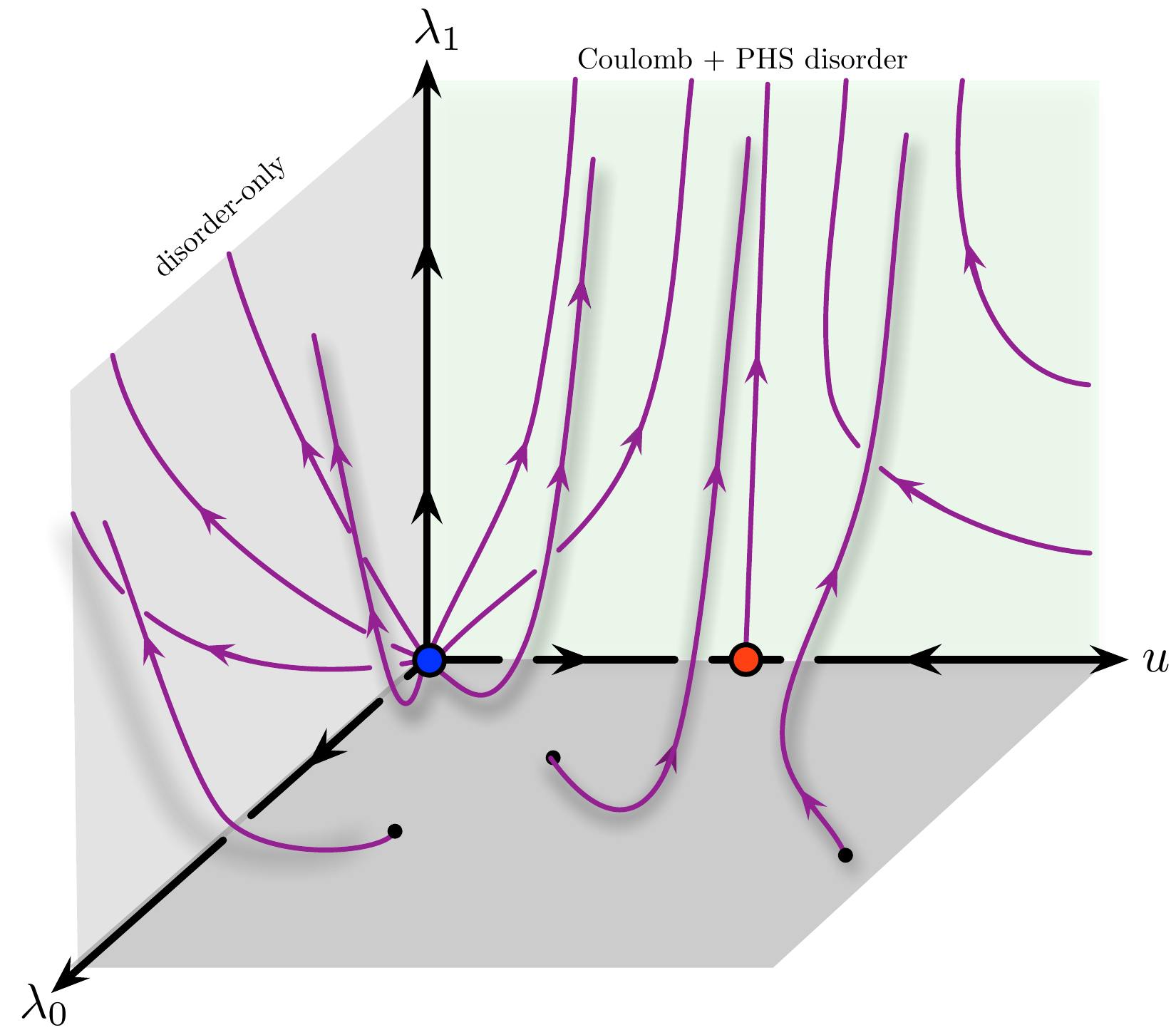}
\caption{\label{3dflow} RG flows in the three dimensional $\lambda_0, \lambda_1, u$ space (recall that $\lambda_i = \frac{2m^2W_i}{\pi^2}$ and $u = \frac{me^2}{8 \pi^2 c}$ are the rescaled couplings.) When $u=0$ the flow is governed by the analysis of the non-interacting disordered problem in Sec.\ref{puredisorder}, and is confined to the $u=0$ plane since there is no way to generate a long range interaction starting from purely short range disorder. In the $u=0$ plane there is only one (Gaussian) fixed point (blue dot), and this is unstable. The flow is to strong disorder. Meanwhile, on the $\lambda_0=\lambda_1=0$ line the flow is given by the clean system analysis of \cite{Abrikosov}. The Gaussian fixed point is unstable and there is a stable fixed point at $u=O(\epsilon)$ (red dot). However, this is unstable to turning on disorder, with $\lambda_1$ being a relevant direction. In the $\lambda_0=0$ plane, the system has exact particle-hole symmetry (PHS), and the flow is confined to this plane because a non-zero $\lambda_0$ would break particle-hole symmetry (and hence cannot be generated by loop corrections if absent in the bare theory); the flow is to strong, purely vector disorder. Finally, when all three couplings are non-zero the flow is also to strong disorder, but this time to strong $\lambda_0$ {\it and} $\lambda_1$. Note that even if $\lambda_1=0$ in the bare theory, non-zero $\lambda_0$ generates non-zero $\lambda_1$, so there are no flows confined to the  plane of pure scalar disorder and interactions.}
\end{figure}

These equations do not have any fixed points at non-zero disorder. To see this, note that the $\beta$-function for $\lambda_1$ is strictly positive, so $\lambda_1$ inevitably grows under RG. Even if we start with $\lambda_1 = 0$, this term is generated by $\lambda_0$, so in the presence of disorder the RG  inevitably flows to large $\lambda_1$. This in turn makes the $\beta$-functions for $u$ and $\lambda_0$ positive, driving a growth in these parameters as well.  Thus, there are no finite-disorder fixed points, and the problem flows to strong disorder {\it and} strong interactions whereupon the perturbative RG is no longer controlled.

We now turn to the perturbative stability of the zero-disorder Abrikosov fixed point against the inclusion of a small amount of quenched  disorder. The relevance of vector disorder is controlled by a  strictly positive $\beta$ function, so the Abrikosov fixed point is manifestly unstable to its addition. Meanwhile, the scaling dimension of scalar disorder  at the Abrikosov fixed point is $- \epsilon$ i.e. pure scalar disorder is {\it irrelevant} at the Abrikosov fixed point. However, it 
inevitably produces vector disorder that grows under the RG and ultimately destabilizes the phase. Thus, introducing any disorder ultimately drives the flow away from the Abrikosov fixed point to strong coupling; the Abrikosov fixed point is thus {\it unstable} to disorder (see Fig.\ref{3dflow}). We thus conclude that quadratic band crossings with disorder and interactions inevitably flow to strong coupling where the perturbative RG analysis is no longer controlled: arbitrarily weak disorder destroys the putative non-Fermi liquid phase. 

In a sense, this is as far as the RG can take us, as long as we restrict ourselves to strictly perturbatively valid statements. However, we may push the  RG one step further and analyze the behavior of various couplings as we exit the perturbative regime en route to strong coupling, and attempt to make some sense of the strongly disordered interacting theory.

 \section{\label{sec:strongflow} Strong Coupling Trajectories}
 
 We now turn to an analysis of the manner in which the system flows to its strong-coupling limit, and out of the regime of the perturbative RG employed here. From (\ref{flowequations2}) we find that $\lambda_1$ has a strictly positive $\beta$-function, i.e. it is monotonically increasing under the RG. Therefore, we may view this as an `RG time'; reparametrizing the flow in terms of $\lambda_1$ reduces the problem to a pair of flow equations for $u, \lambda_0$: from (\ref{flowequations1}) and (\ref{flowequations3}) we have
 \begin{eqnarray}
 \frac{d\lambda_0}{d\lambda_1}  
&=&   \frac{dl}{d\lambda_1}{ \lambda_0 \left[\epsilon + 2 \lambda_0 + 2(N+1) \lambda_1 - \frac{60 N_f + 16}{15} u \right] },\nonumber\\
 \frac{du}{d\lambda_1} 
 &=&  \frac{dl}{d\lambda_1} {u\left[ \epsilon - \frac{30 N_f + 8}{15} u + \frac32 (\lambda_0 + N \lambda_1) \right]},
\end{eqnarray}
where $\frac{dl}{d\lambda_1}$ may be determined from (\ref{flowequations2}). Next, we observe that $\epsilon/\lambda_1 \rightarrow 0$ under the RG. Therefore, along the flow to strong coupling this `tree level' term is eventually unimportant, and we can simply look at the flow of ratios of couplings, viz. $x = \lambda_0/\lambda_1$ and $y = u/\lambda_1$.  Setting $\epsilon/\lambda_1$ to zero and rewriting we obtain \cite{Vafek} 
  \begin{align}
 \frac{dx}{d\ln\lambda_1}   &= x \left[-1  +  \frac{ 2 x + 2(N+1) - \frac{60 N_f + 16}{15} y  }{\frac{2x}{N} + \frac{2 N^2-N+2}{N} + 2 y \left(\frac{N-1}{N} - \frac{8}{15} \right)+ \frac{x^2}{N}}\right]\label{xstability}\\
 \frac{dy}{d\ln\lambda_1}  &= y \left[-1 + \frac{ - \frac{30 N_f + 8}{15} y + \frac32 (x + N)}{\frac{2x}{N} + \frac{2 N^2-N+2}{N} + 2 y \left(\frac{N-1}{N} - \frac{8}{15} \right) + \frac{x^2}{N}}\right]. \label{ystability}
\end{align}

Fixed points of these equations represent `fixed trajectories' of the flow to strong coupling. Setting $N=5$ we find that the fixed trajectories are determined by the simultaneous equations
  \begin{eqnarray}
0 &=& x \left[-1  +  \frac{ \left[ 2 x + 12 - \frac{60 N_f + 16}{15} y \right] }{ \left[ x \frac{2}{5} + \frac{47}{5} +  y \frac{8}{15} \right] + \frac15 x^2}\right] \label{xeq}\\
0 &=& y \left[-1 + \frac{\left[  - \frac{30 N_f + 8}{15} y + \frac32 (x + 5) \right]}{ \left[ x \frac{2}{5} + \frac{47}{5} +  y \frac{8}{15}\right] + \frac15 x^2}\right] \label{yeq}
\end{eqnarray}
with the additional constraint that $(x,y)$ are real and non-negative, corresponding to physical solutions accessible via the flow equations.

One obvious solution is $(x_*, y_*) = (0,0)$, but this is unstable: the linearized flow equations in the vicinity of this fixed point are
\begin{eqnarray}
\left.\frac{d}{d\ln\lambda_1}\right|_{(x_*,y_*) =(0,0)}\left(\begin{array}{c} \delta x \\ \delta_y \end{array}\right) \approx \left(\begin{array}{cc} \frac{13}{47}& 0\\0&-\frac{19}{94}\end{array}\right)\left(\begin{array}{c} \delta x \\\delta y\end{array} \right),\,\,\,\,\,\,\,\,\,
\end{eqnarray}
so that it is stable against non-zero $y$ but unstable against non-zero $x$. This is the fixed trajectory corresponding to flows in the $(u, \lambda_1)$ plane: both disorder and interactions increase in strength but the theory is asymptotically dominated by a strong-coupling fixed point where vector disorder dominates the interactions. As noted previously, enforcing particle-hole symmetry permits us to set scalar disorder to zero (i.e. take  $x=0$); breaking this symmetry inevitably introduces some $x>0$, upon which we flow to a different fixed trajectory, that we now determine.

We next seek solutions where only one out of $x$ and $y$ is non-zero. There are no solutions with $x=0$ and $y>0$. However, we find a solution with $y_*=0$ and $x_*  = 4+\sqrt{29}\approx 9.38$.  We have already determined that $x$ is a relevant perturbation along the $y=0$ line; from inspection of (\ref{xstability}) that the flow for large $x$ is to smaller $x$. As there are no other fixed points along this line, we conclude that this fixed point is stable along $x$. The stability to changes in $y$ may be determined straightforwardly: we find that
\be
\left. \frac{d \delta y}{d\ln\lambda_1}\right|_{{(x_*,y_*) \approx(9.38,0)}} \approx  - 0.30 \delta y,
\ee
indicating that $y$ flows back to zero, so that this fixed point is stable to changes in both $x$ and $y$. This corresponds to an RG trajectory along which {\it both} vector and scalar disorder grow; while interactions grow slower and are hence dominated by disorder, the {\it ratio} of scalar and vector disorder flows to a fixed value of 9.38. 

Finally, we find that there are no solutions where both $x$ and $y$ are positive, with $N_f$ a positive integer: there are no fixed trajectories along which interactions grow as quickly as disorder.

\begin{figure}
\includegraphics[width=.8\columnwidth]{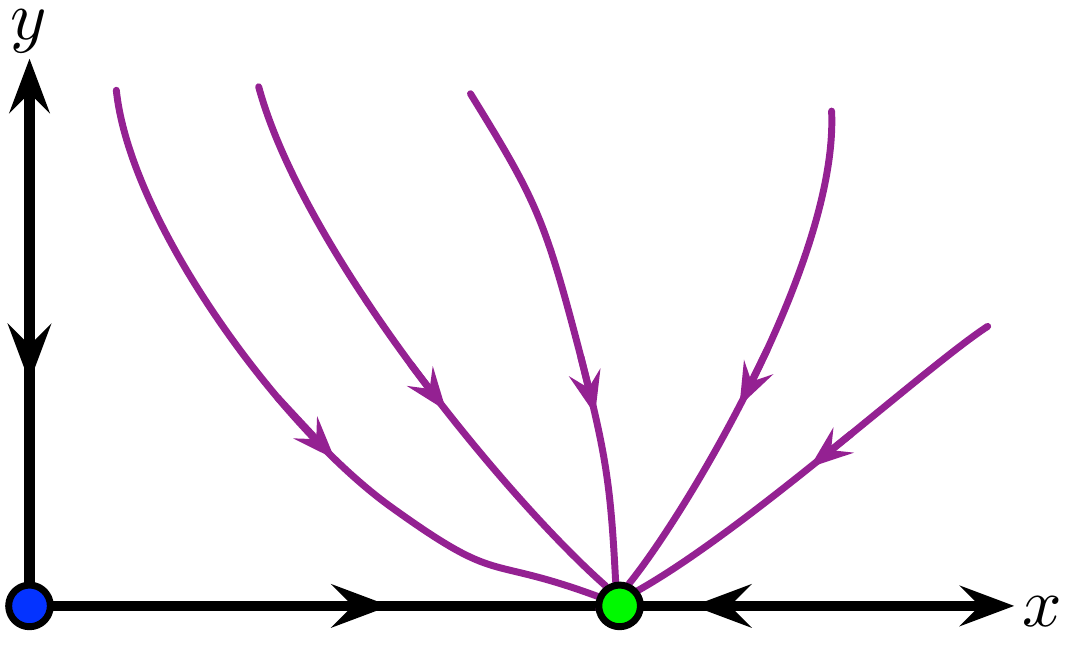}
\caption{Flow of the ratios $x = \lambda_0/\lambda_1$ and $y = u/\lambda_1$ as the system flows to strong coupling, obtained by extrapolation from the  perturbative RG equations at one loop. Note that $y$ flows to zero so the theory is asymptotically non-interacting. Exact  particle-hole symmetry ($x=0$) is preserved by the RG, so that we stay at $x=0$ if the bare theory has this symmetry; otherwise the system flows to a particular ratio $\lambda_0/\lambda_1 \approx 9.38$. \label{trajectoryflow}}
\end{figure}

The resulting flow diagram is sketched in Fig.\ref{trajectoryflow}. In both cases --- with or without the imposition of particle-hole symmetry --- the problem flows to an effectively {\it non-interacting}, strongly-disordered theory. The physics in this limit may therefore be accessible within a {\it non-interacting} sigma model description --- an enormous simplification with respect to the long-range interacting model (\ref{fullactionMoon}) with which we began. We find an additional simplification emerges en route to strong coupling: the fixed-point trajectories are be characterized by a fixed ratio of scalar to vector disorder ($\lambda_0/\lambda_1\approx 9.38$ within the perturbative theory, though the precise value is likely modified). Thus, we conjecture that the strong-coupling behavior is controlled by a single fixed point where both scalar and vector disorder are strong and scale similarly.

\section{\label{sec:conclusions}Concluding Remarks}

We have found that a system hosting a quadratic band crossing, whose microscopic Hamiltonian contains Coulomb interactions and short-range-correlated disorder flows, on long lengthscales, to a strongly coupled phase that lies outside the perturbative renormalization group framework. An analysis of this flow to strong coupling  suggests that disorder grows asymptotically faster than interactions, so that the strong coupling phase  likely can be described within a {\it non-interacting} sigma model. This, in turn, suggests that the complex long range interacting problem maps (at least as far as its low energy properties are concerned) to a non-interacting problem whose solution may be found in the extensive literature on Anderson insulators. Of course, insofar as this conclusion is based upon extrapolating the one loop flow to strong coupling, it is potentially suspect, and these conclusions should in principle be verified through an {\it interacting} sigma model calculation. Crucially, however, our one-loop results do suggest that the sigma model need not treat disorder and Coulomb interactions on an equal footing, rather it should be adequate to begin with a disorder-only sigma model, add the Coulomb interaction as a weak perturbation, and examine the relevance or irrelevance thereof.

The specific disorder-only non-interacting sigma model that is a candidate to describe our systems depends on the particular symmetries at hand. In the general case, when scalar potential disorder (that breaks particle-hole symmetry) is present, the only remaining anti-unitary symmetry is time reversal; as this  squares to $-1$, the problem falls into the three-dimensional symplectic class (Hamiltonian class AII in the Altland-Zirnbauer classification~\cite{EversMirlin}). Note that the ratio of scalar to vector disorder flows to a constant value $4 + \sqrt{29} \approx 9.38$, so even though the sigma model will contain both vector and scalar disorder, there will be only one independent `stiffness' parameter. In principle, the sigma model can also contain a $Z_2$ `topological' term~\cite{Schnyderclassification}. There should thus be two distinct possible behaviors at strong coupling, depending on whether or not the system is topologically non-trivial (a distinction that is immaterial for our perturbative analysis, but which should become important at strong coupling). The topologically trivial sigma model is believed to support both a diffusive metallic phase, and a localized phase, with a localization transition that may be described using methods outlined in e.g. Ref.~\onlinecite{EversMirlin}. The topologically non-trivial sigma model may support a still richer behavior, but has not been solved as far as we are aware. As the bare Hamiltonian contains Coulomb interactions, our analysis thus presents an intriguing scenario: a rare example where a (zero temperature) many body localization transition with long range interactions may be analytically tractable, because the system flows to an effectively non-interacting description.
We also flag  the possible role of {\it statistical} particle-hole symmetry in modifying details of such a sigma-model analysis.

If {\it exact} particle-hole symmetry is imposed on the problem --- by setting $W_0=0$ in the bare theory ---  the flow to strong coupling is described asymptotically by a flow to strong vector potential disorder alone (i.e.,$x=0$). The resulting theory has both $C^2 = -1$ and $T^2 = -1$, and corresponds to the chiral symplectic class \cite{tenfold} (Hamiltonian class CII in the Altland-Zirnbauer classification) in three dimensions. Again there will be a localized and a delocalized phase, with the transition now being governed by the sigma model appropriate to the chiral symplectic class. Additionally, the localized phase may itself have anomalous features, such as a singular low energy density of states \cite{Wegnerclassification} (note that this is forbidden on general grounds in the absence of particle-hole symmetry \cite{Wegner}). As in the $W_0\neq 0$ case, there is the possibility of realizing an analytically tractable zero temperature {\it many-body localization} transition, for a three dimensional system with long range interactions, but now in a different universality class. Finally, this symmetry class too can support a $Z_2$ topological term \cite{Schnyderclassification}, that may produce still richer behavior if present.

 Again, we 
 note that a more careful analysis would examine the role of interactions in the vicinity of the putative sigma-model critical point.
However, the results of the perturbative RG do offer hope that such a procedure may indeed yield an  interesting and analytically tractable example of an MBL transition; exploring this  would be a worthwhile challenge for future work. More prosaically, it could provide a valuable toy model for the three dimensional electron glass \cite{Ovadyahu}, a system whose analytical understanding remains poor to date. All these facts serve as additional motivation for for performing the sigma model calculation, that we defer to future work.

One issue we have not thus far addressed is that of rare-region (`Griffiths') effects, that are known to dominate the low energy behavior for non-interacting Weyl semimetals with scalar disorder~\cite{NandkishoreHuseSondhi, pixleyhuse}. Since in the present case the problem flows to strong disorder already at the perturbative level, we might conjecture that Griffiths effects will be less important than in the Weyl semimetal~\cite{NandkishoreHuseSondhi}. However, Griffiths effects also dominate the low-energy behavior for two dimensional particle-hole symmetric localization~\cite{MotrunichDamleHuse}. The relevance of such effects for our problem remains an open question.

We now discuss one further subtlety. We assumed in the discussion above that a localized phase could be realized at strong disorder. However, it has been argued that localization is incompatible with Coulomb interactions because they are `too long ranged'~\cite{Burin}. Separately, it has also been argued that MBL is incompatible with particle-hole symmetry~\cite{AIIIMBL}, a fact related to its non-Abelian structure. Systems in dimensions greater than one have also been argued to be non-localizable \cite{avalanches}. However, the arguments of Refs.~\onlinecite{Burin, AIIIMBL, avalanches} all consider systems at finite energy density (non-zero temperature), and hence do not apply directly to the zero-temperature scenario that we consider here. It could be interesting to consider turning on a small but non-zero temperature, and investigating thermalization in the resulting system. Refs.~\onlinecite{Burin, AIIIMBL, avalanches} suggest that localization in the strongly disordered phase should be instantly destroyed upon turning on a non-zero temperature. However, these are asymptotic statements valid only beyond  characteristic length and time scales, below which the system looks localized, and that must diverge as $T \rightarrow 0$. An investigation of these scales and how they diverge (similar to Ref.~\onlinecite{QHMBL}) would also be an interesting problem for future work. The critical point itself presumably becomes an avoided critical point at non-zero temperature, and again, this avoided criticality is likely worthy of further study. 

Finally, in deriving the symmetry-constrained action we made two simplifications, in assuming that (a) anisotropy terms could be ignored and (b) exact time reversal symmetry could be imposed. It would be interesting to relax these assumptions and explore the resulting physics. This too we leave to future work. 

We conclude by discussing other settings to which calculations analogous to those reported here may apply. The fundamental property of the problem  is that long-range interactions and disorder are both relevant, with the same scaling dimension, and thus must be treated on an equal footing. Let us consider a general problem with dynamic exponent $z$, Fermi surface co-dimension $d$, and a long range interaction that obeys Gauss's law in $D$ spatial dimensions. Note that we require that $D \ge d$ for the problem to be physically sensible. The interaction potential will then fall off as $1/r^{D-2}$. We can verify that disorder will have tree-level scaling dimension $2z-d$, whereas the long-range interaction will have scaling dimension $z - D+2$; the two are equally relevant if these scaling dimensions coincide, i.e., if
\begin{equation}
z = d - D + 2. \label{equality}
\end{equation}
[Note that as properties of the clean, non-interacting theory, $z, d, D$ must all be positive integers]. The scaling dimension is $2z-d$, and the perturbation theory is developed about $d=2z$. The quadratic semimetal studied here corresponds to $D = d$ and $z=2$, and the perturbation theory was developed about $d = 4-\epsilon$. One could also consider $d = D-1$ and $z=1$, and develop the perturbation theory about $d=2$. This situation describes graphene (where the Fermi surface co-dimension is two but the Coulomb interaction lives in three dimensions), and was analyzed in Refs.~\onlinecite{Stauber, YeSachdev, HerbutVafek}. However, it also describes three-dimensional systems with line nodes and $1/r$ interaction, as well as two-dimensional systems with a Fermi surface (co-dimension 1) and $\log(r)$ interactions. In these settings the interplay of disorder and interactions has not been explored, and it would be interesting to do so in the future. 

\begin{acknowledgements}
We thank Subir Sachdev for a detailed and illuminating discussion regarding various aspects of the calculation, and Victor Gurarie for useful discussions. We also acknowledge support from NSF Grant DMR-1455366 (SAP).
\end{acknowledgements}

\begin{appendix}
\section{\label{sec:CleanRGOnlyZS} Absence  of VC, BCS, ZS' contributions in the clean System}
In this appendix, we show that the VC, BCS, and ZS' diagrams do not contribute to the clean-system RG flow. Since (as we demonstrate) these diagrams make vanishing contribution we will not be precise about combinatorial factors, but  simply focus on showing that there is no logarithmic divergence in these channels. The VC diagram with two Coulomb lines cannot make a contribution, since the residue is not renormalized by the Coulomb interaction and there is a Ward identity on the product of vertex and residue. Explicitly, the vertex correction {\it from} Coulomb lines {\it to} Coulomb lines has a relative minus sign compared to ZS and takes the form 
\begin{align}
\Gamma^{\text{VC}}_{00} &\propto - \frac{e^4}{c^2q^{2}}\int d\omega d^4 p \frac{G(\omega, \vec{p}) G(\omega, \vec{p}+\vec{q})}{p^{2}}.
\end{align} 
After noting that terms odd in frequency vanish upon integrating over frequency and canceling terms that vanish upon angular integration over $\vec{p}$ we have 
\begin{eqnarray}
\Gamma^{\text{VC}}_{00}&\propto& 2 \frac{e^4}{q^{2}} \int d \omega d p^4 \frac{-\omega^2 + \sum_{\alpha} d_{\alpha}(\vec{p}) d_{\alpha}(\vec{p+q})}{p^{2}(\omega^2 + p^4)(\omega^2 + |\vec{p}+\vec{q}|^4)} \nonumber\\ &\approx&- \frac{e^4}{2 q^{2}} \int_{\infty}^{-\infty} d \omega \int_0^{\infty} d p^4 \frac{-\omega^2 + p^4}{p^{2}(\omega^2 + p^4)^2}
\end{eqnarray}
where in the last step we have expanded the integral to leading order in small $q$, with higher order terms being less relevant (producing a short range interaction rather than a long range Coulomb interaction). Now the frequency integral (which can be done exactly by the method of residues) vanishes, such that the Coulomb correction to the Coulomb vertex is zero, consistent with what we would expect from the Ward identity. 

We turn next to the ZS' and BCS diagrams. These correspond to ladder and twisted ladder diagram topologies (``Cooperon'' and ``diffuson''). It is important when evaluating these to keep track of the total momentum transfer \cite{NandkishoreLevitov}. Let the incoming momenta be $\vec{k_1}$ and $\vec{k_2}$, and the outgoing momenta be $\vec{k_1} + \vec{q}$ and $\vec{k_2} - \vec{q}$. The external momenta $k_1$ and $k_2$ can be set to zero, but the momentum transfer $\vec{q}$ cannot --- it is needed to split a high order pole in the integrand coming from the doubled Coulomb line.  The sum of these two diagrams with two Coulomb lines attached (and $\vec{q}$ kept track of) yields (note the overall minus sign with respect to ZS)
\begin{eqnarray}
\Pi^{\text{ZS'}}_{00} + \Pi^{\text{BCS}}_{00} &\propto&  - e^4\int \frac{d \omega d^4 p}{p^{2} |\vec{p}-\vec{q}|^{2}} G(\omega, \vec{p}) \left[G(\omega, \vec{p} - \vec{q})\right.\nonumber\\& & \left.+ G(-\omega, - \vec{p})\right] \\ &=&  - e^4\int \frac{d \omega d^4 p}{p^{2} |\vec{p}-\vec{q}|^{2}} \frac{d(\vec{p}) \cdot d(\vec{p-q}) + |\vec{p}-\vec{q}|^4}{(\omega^2 + p^4)(\omega^2 + \vec{p}-\vec{q}|^4)}\nonumber
\end{eqnarray}
where we have simplified, used $d(-\vec{q}) = d(\vec{q})$, and  dropped terms that vanish upon frequency or angular integration. The integral over frequency may then be performed and yields 
\begin{equation}
\Pi^{\text{ZS'}}_{00} + \Pi^{\text{BCS}}_{00}  \propto -  \frac{e^4}{2} \int d^4 p\frac{d(\vec{p}) \cdot d(\vec{p-q}) + |\vec{p}-\vec{q}|^4}{p^4 |\vec{p}-\vec{q}|^4(p^{2} + |\vec{p}-\vec{q}|^{2})}.\label{divergingBCSZSp}
\end{equation}
We now see that we cannot simply set $q$ to zero because the  integral in (\ref{divergingBCSZSp}) diverges as $1/q^{2}$ i.e. it produces a correction to the Coulomb line. Formally, this diagram diverges logarithmically, but this divergence originates in momenta $p \ll q$. Since we are performing an RG calculation, we should assume external momenta are much smaller than internal momenta, i.e. take $q \ll p$. In that case we can evaluate the integral by Taylor expanding the integrand in small $q$ to get 
\begin{equation}
\Pi^{\text{ZS'}}_{00} + \Pi^{\text{BCS}}_{00} \propto -  \frac{e^4}{2} \int d^4 p \frac{1}{p^8} \frac{2 p^4}{2p^2} \sim \int \frac{d^4 p}{p^6} 
\end{equation}
where $q$ provides an infrared cutoff on the integral, and we have dropped terms at higher orders in $q$ which are less relevant. 
This then yields an integral that scales as $\frac{1}{q^2}$ with no logarithmic divergence. Thus, done correctly, the ZS' and BCS diagrams produce a {\it constant} correction of the Coulomb line, but not a log divergent contribution, and thus they do not contribute to the $\beta$ functions. 

\end{appendix}

\bibliography{LAB_bib}

\end{document}